\documentclass[a4paper,11pt]{article}
\pdfoutput=1 

\usepackage{jheppub} 
\usepackage{hyperref}
\usepackage{caption}
\usepackage{subcaption}
\usepackage[T1]{fontenc} 
\usepackage{listings}
\usepackage{xcolor}
\usepackage{slashed} 
\usepackage{multirow}
\usepackage{dsfont}
\usepackage{array}
\usepackage{orcidlink}
\allowdisplaybreaks

\def\eps{\epsilon}

\newcommand{\hexagonLS}{{\Delta_6}}

\newcommand{\+}{{\dt+}}

\title{A computation of 
two--loop six--point Feynman integrals in 
dimensional regularization 
}
\author[a]{Johannes Henn}
\author[a]{Antonela Matijašić}
\author[a]{Julian Miczajka}
\author[b]{Tiziano Peraro}
\author[c]{Yingxuan Xu}
\author[d,e]{Yang Zhang\orcidlink{0000-0001-9151-8486}}

\affiliation[a]{Max--Planck--Institut f\"{u}r Physik, Werner--Heisenberg--Institut, D--85748 Garching bei M\"{u}nchen, Germany}
\affiliation[b]{Dipartimento di Fisica e Astronomia, Universitá di Bologna e INFN, Sezione di Bologna, via
Irnerio 46, I-40126 Bologna, Italy}
\affiliation[c]{Humboldt--Universit\"at zu Berlin, Institut f\"ur Physik, Newtonstraße 15, 12489 Berlin, Germany}
\affiliation[d]{Interdisciplinary Center for Theoretical Study, University of Science and Technology of China,\\ Hefei, Anhui 230026, China}
\affiliation[e]{Peng Huanwu Center for Fundamental Theory, Hefei, Anhui 230026, China}

\emailAdd{henn@mpp.mpg.de,
amatijas@mpp.mpg.de,
miczajka@mpp.mpg.de,
tiziano.peraro@unibo.it,
yingxu@physik.hu-berlin.de,
yzhphy@ustc.edu.cn
}

\preprint{MPP-2024-53, USTC-ICTS/PCFT-24-11}

\abstract{
We compute three families of two--loop six--point massless Feynman integrals in dimensional regularization, namely the double--box, the pentagon--triangle, and the hegaxon--bubble family.
This constitutes the first analytic computation of two--loop master integrals with eight scales. We use the method of canonical differential equations. We describe the corresponding integral basis with uniform transcendentality, the relevant function alphabet, and analytic boundary values at a particular point in the Euclidean region up to the fourth order in the regularization parameter~$\epsilon$. The results are expressed as one--fold integrals over classical polylogarithms suitable for fast and high--precision evaluation.
}

\begin{document}

\maketitle


\section{Introduction}
With the data accumulated in the runs of LHC, multiple jet production processes become more and more interesting~\cite{ATLAS:2015xtc}. A precise theoretical prediction for those processes requires the next--to--leading order (NLO) and next--to--next--to--leading order (NNLO) order computation of multi--leg Feynman integrals and scattering amplitudes.
The last few years have seen substantial advancements in the computation of two--loop Feynman integrals, scattering amplitudes, and cross sections. Analytical calculations for two--loop five--point massless integrals have been achieved  \cite{Gehrmann:2018yef,Abreu:2018rcw,Abreu:2018aqd,Chicherin:2018mue,Chicherin:2018old}, providing a comprehensive set for  $2\rightarrow 3$ massless scattering process. 
Notably, a dedicated computer implementation has been established for efficient and reliable evaluation of five-point Feynman integrals in the physical region, detailed in Ref.~\cite{Chicherin:2020oor}. The analytical progress has already found applications in various amplitudes computations \cite{Gehrmann:2015bfy,Badger:2017jhb,Badger:2018enw,Abreu:2017hqn,Chawdhry:2020for,Abreu:2020cwb,Abreu:2018zmy,Badger:2019djh,Abreu:2021oya,Agarwal:2023suw,Chawdhry:2021mkw,Agarwal:2021vdh,DeLaurentis:2023izi,DeLaurentis:2023nss,Badger:2021imn} and phenomenological processes~\cite{Kallweit:2020gcp,Czakon:2021mjy,Chawdhry:2021hkp,Badger:2023mgf,Badger:2021ohm}. 
In addition, all master integrals for two--loop five--point scattering with one off-shell leg~\cite{Abreu:2020jxa, Canko:2020ylt, Abreu:2021smk, Abreu:2023rco} were computed analytically and the results were further optimized for fast evaluation of the functions in the physical scattering region \cite{Chicherin:2021dyp}. These integrals play a crucial role in NNLO scattering amplitudes for two--jet--associated W--boson/Z--boson production in QCD~\cite{Badger:2021nhg}. Recently, the two-loop five-point pentagon-box integral configuration with one internal massive propagator was also evaluated~\cite{Badger:2022hno}, contributing to top-quark pair production with a jet at hadron colliders.


On the other hand, for the production of four or more jets, only NLO cross sections are known in the literature 
\cite{Bern:2011ep,Badger:2013yda}. On the amplitude side, one-loop six-gluon \cite{Ellis:2006ss,Dunbar:2008zz} and photon amplitudes \cite{PhysRevD.49.2197,Ossola:2007bb,Binoth:2007ca,Bernicot:2007hs} are known, but for the two-loop six-gluon amplitudes, only the special all-plus case is presented in the literature~\cite{Dunbar:2016gjb,Dunbar:2017nfy}. Almost nothing is known about the two--loop six--particle function space, even for planar massless scattering. The lack of known Feynman integrals represents one of the main obstructions on the way toward NNLO results.

Beyond these phenomenological motivations, it is also interesting to study six--point Feynman integrals from a more theoretical perspective. 
In planar $\mathcal{N}=4$ super Yang-Mills (sYM), the function space of planar six--particle scattering is in fact conjectured to be known to all loop orders \cite{Golden:2013xva}. This conjecture, together with other physical insights such as behavior in limits and analytic properties, has been used to bootstrap six--particle amplitudes to staggering loop orders, see e.g. \cite{Caron-Huot:2016owq,Dixon:2023kop}.

Thanks to dual conformal conformal symmetry enjoyed by that theory \cite{Drummond:2008vq}, the number of independent dimensionless variables is three, as opposed to seven in the QCD case (for four--dimensional external states), so the situation is rather different. However, higher--point amplitudes in sYM have been observed to have similarities with lower-point QCD amplitudes \cite{Chicherin:2020umh,Bossinger:2022eiy}. In particular, the eight--point sYM case, described by an infinite--dimensional cluster algebra~\cite{Arkani-Hamed:2012zlh,Golden:2013xva}, might be related to the six--point case studied here.
Several recent papers focus on understanding the symbol alphabet of the eight--particle sYM amplitudes \cite{Arkani-Hamed:2019rds,Herderschee:2021dez,Ren:2021ztg,Henke:2021ity,He:2021non,Yang:2022gko,He:2022ujv}.

Related to this, considerable recent work is dedicated to understanding the locus of singularities of Feynman integrals (and hence, scattering amplitudes). This is closely related to the question of the relevant function space and alphabet of iterated integrals. Recent work includes the study of Landau singularities \cite{Fevola:2023kaw,Fevola:2023fzn,Helmer:2024wax}, as well as different methods to directly determine the alphabets~\cite{He:2023umf,Jiang:2024eaj}.

Knowing the full planar six--particle alphabet would open the door to potential bootstrap application. For example, the Wilson loop with Lagrangian insertion in 
sYM~\cite{Alday:2011ga} was observed to be related to all--plus amplitudes in pure Yang--Mills theory~\cite{Chicherin:2022bov,Chicherin:2022zxo}, and hence depends on the same function space.

Having access to analytic results for two--loop six--point Feynman would greatly help with those questions.
Hence, our long--term goal is to calculate all two--loop six--point massless integrals. 
In this paper, we take 
a step towards that goal, by providing canonical differential equations for three out of the six planar families that are required for describing any six--point amplitude in any theory. Eventually, we would compile our result in an automatic package for fast and efficient computation of these integrals, like the five--point integral package~\cite{Chicherin:2020oor}.

In the Ref.~\cite{Henn:2021cyv}, the potential of computing two--loop six--point integrals in dimensional regularization via integrals with uniform transcendental (UT) weights and canonical differential equation (DE) \cite{Henn:2013pwa} was demonstrated. The maximal cuts of a UT basis for genuine (non--factorizable) two--loop six--point massless planar integrals were found and the diagonal blocks of the canonical differential equations and symbol letters were calculated. On the other hand, one--loop six--point massless integrals were calculated in \cite{Henn:2022ydo} up to transcendental weight four. 

In this paper, we compute three families of genuine two--loop six--point massless Feynman integrals beyond the maximal cut: the double--box, pentagon--triangle and hexagon--bubble families. Since these families are subtopologies of the remaining three families, namely the pentagon-box, double-pentagon and hexagon-box, it is necessary extend their definition beyond the maximal cut first. We find the complete integral UT basis for the three families, with the Baikov representation \cite{Baikov:1996cd,Dlapa:2021qsl,Chen:2022lzr} and Gram determinant construction. 

The corresponding canonical differential equations are derived via finite--field techniques \cite{vonManteuffel:2014ixa,Peraro:2016wsq}, where an a--priori knowledge of the symbol alphabet is tantamount to efficiently determine the differential equations. To that end, we employed an efficient method to find algebraic symbol letters with information of square roots in the UT integral definition and known even letters~\cite{Matijasic:2024gkz}. The boundary values at a specific point are obtained analytically based on the singularity structure of the canonical differential equations in the Euclidean region. Furthermore, we set up integration paths in the Euclidean region to solve the canonical differential equations. The weight one and two parts of those integrals are presented in terms of classical polylogarithms, while weight three and four parts are presented as one--fold iterated integrals suitable for fast numerical evaluation. 

Throughout this paper, we use a scheme such that external momenta are in four dimensions, while the loop integration is $(4-2\eps)$-dimensional. This is compatible with both {`t~Hooft}--Veltman as well as four--dimensional helicity scheme at the level of the amplitudes. One advantage for computations in this scheme is that we can use a momentum twistor parametrization \cite{Hodges:2009hk} to rationalize a lot of the square roots in the canonical differential equation. Furthermore, in this convention, the number of irreducible scalar products is smaller than the corresponding number for D-dimensional external momenta. (See  section~\ref{ref:subsection_Integral_families} for the discussion.) There are eight scales in the computation and our work constitutes the first two--loop eight--scale master integral computation via canonical differential equation.


This paper is organized as follows: In section 2, we set up the six--point kinematics and conventions. 
In section 3, we describe how to efficiently calculate the differential equation by fitting against an ansatz in terms of the hexagon alphabet and we briefly introduce our method for finding odd letters.
In section 4, we present the complete UT basis for the two--loop six--point double box, pentagon triangle and hexagon bubble families. In section 5, we identify all symbol letters that show up in the canonical differential equations for those three families, defining the symbol alphabet. 
In section 6, we perform the integration over paths in the Euclidean region to solve the Feynman integrals in the three families up to weight four. In section 7, we describe all the auxiliary files we provide. Finally, in section 8, we summarize this paper and provide an outlook for future directions.

\section{Kinematics and Conventions}
\subsection{Six--particle kinematics and Lorentz invariants}
Consider a generic scattering process for six massless particles which are all considered as outgoing,
\begin{gather}
0\rightarrow 1+2+ \dots +6.
\end{gather}
The six external momenta labeled as $p_i$'s, $i=1,\dots,6$ are constrained by momentum conservation and on-shell-ness,~i.e.
\begin{equation}
     \sum_{i=1}^6 p_i=0,\quad\text{and}\quad p_i^2=0, \quad i=1,\ldots,6.
\end{equation}
The corresponding Mandelstam variables are
\begin{gather}
    s_{ij}=(p_i+p_j)^2, \quad s_{ijk}=(p_i+p_j+p_k)^2, \quad 1\leq i,j,k \leq 6.
\end{gather}
Moreover, the Gram determinant is defined as
\begin{gather}
    G\left(
\begin{array}{ccc}
u_1 & \dots & {u_n} \\
v_1 & \dots & {v_n} \\
\end{array}
\right)=\mathrm{det}(2 u_i\cdot v_j).
\end{gather}
Here the right--hand side is the determinant of the $n\times n$ matrix with the entries $2(u_i\cdot v_j),1\leq i,j \leq n$. We introduce the abbreviation $G(i_1,\dots,i_k)$ defined as
\begin{gather}
    G(i_1,\dots,i_k)\equiv     G\left(
\begin{array}{ccc}
p_{i_1} & \dots & p_{i_k} \\
p_{i_1} & \dots & p_{i_k} \\
\end{array}
\right).
\end{gather}
For a scattering process of $N$ particles, there are $3N-10$ independent parity--even Lorentz invariants, if external momenta are confined to four dimensions. However, in general dimensions, there are $\frac{N(N+1)}{2}$ scalar products, of which $\frac{N(N-3)}{2}$ are independent. While $3N-10$ aligns with $\frac{N(N-3)}{2}$ for four--point and five--point scattering, the number of variables becomes different in six--point scattering. 
To be explicit, there are nine independent Mandelstam variables for $D$--dimensional external momenta
\begin{equation}
    \vec{s}=\{s_{12},s_{23},s_{34},s_{45},s_{56},s_{61},s_{123},s_{234},s_{345}\}.
\end{equation}
Since only four vectors can be independent in four dimensions, there is a linear relation among any five momenta. After using momentum conservation to eliminate $p_6$, the remaining constraint is captured by the vanishing of the following Gram determinant,
\begin{gather}
\label{eq:GramConstraint}
     G(1,2,3,4,5)=0.
\end{gather}
This reduces the number of independent Mandelstam variables from nine to eight. 

To fully characterize the kinematics of a scattering process, in addition to the parity--even invariants defined above, one needs to specify the sign of one parity--odd pseudo--scalar invariant~\cite{Eden:1966dnq}. A pseudo--scalar can be formed by contracting the antisymmetric Levi--Civita tensor with any four momenta of the scattering process,~i.e.
\begin{gather}
      \epsilon_{ijkl}\equiv 4 \sqrt{-1} \ \varepsilon_{\mu_1 \mu_2 \mu_3
    \mu_4} p_i^{\mu_1} p_j^{\mu_2}p_k^{\mu_3}p_l^{\mu_4},\quad 1\leq
  i,j,k,l \leq 6.
\end{gather}
Only one of these objects is independent, since the product of any two $\eps_{ijkl}$ may be written in terms of a Gram determinant.
In particular, the square of the pseudo scalars equals
\begin{gather}
      \epsilon_{ijkl}^2 = G(i,j,k,l)\,.
\end{gather}

In this paper, all calculations of six--point integrals are carried out in the Euclidean region, which is defined by
\begin{equation}
    s_k <0, \quad k=1,...,9.
\end{equation}
Throughout this region, all Feynman integrals are manifestly real--valued. Further, kinematic divergences only occur at the boundaries of this region, where any of the Mandelstam invariants go to zero. 
In physical scattering regions, such as that for two-to-four-particle scattering, a subset of the Mandelstam variables takes positive values~\cite{Byckling:1971vca,Henn:2022ydo}. 
Such regions can be reached by analytic continuation, but this is beyond the scope of the present paper.

\paragraph{Spinor--Helicity Variables.} We also use spinor helicity notations,
\begin{equation}
  p_{i\mu} \sigma^\mu_{\alpha \dot \beta}=\lambda_{i\alpha}\tilde
  \lambda_{j\dot \beta}, 
\end{equation}
with the Pauli matrices $\sigma^\mu=(I_{2\times 2}, \sigma^1, \sigma^2,
\sigma^3)$. The spinor products are defined as,
\begin{eqnarray}
  \langle ij \rangle &\equiv \lambda_i ^\alpha \lambda_{j,\alpha }\,,\\ 
  \ [ij]   &\equiv \tilde \lambda_{i ,\dot{\alpha}} \tilde
\lambda_j^{\dot{\alpha }} \,.
\end{eqnarray}
In spinor--helicity variables the sign of the $\epsilon_{ijkl}$ is fixed via 
\begin{align}
\label{eq:214}
  2 [ij]\langle jk \rangle [kl] \langle li\rangle = s_{ij}s_{kl}-s_{ik}s_{jl}+s_{il}s_{jk}+ \eps_{ijkl}.
\end{align}
We also introduce the object
\begin{align}
    \hexagonLS &=\text{tr}(\gamma_5 \slashed{p}_1 \slashed{p}_2\slashed{p}_3\slashed{p}_4\slashed{p}_5\slashed{p}_6)\notag\\
    &= \langle12\rangle[23]\langle 34\rangle[45]\langle56\rangle[61]-[12]\langle 23\rangle[34]\langle45\rangle[56]\langle61\rangle,
\end{align}
which naturally arises as the four--dimensional limit of the leading singularity of the one--loop hexagon integral, c.f.~\cite{Henn:2022ydo}. In the definition of $\Delta_6$ we employ slashed momenta and the chiral $\gamma_5$, both of which are $4\times4$ matrices given by
\begin{equation}
    \slashed{p}_i = \begin{pmatrix}0 & |i]\langle i| \\ |i\rangle [i| & 0
    \end{pmatrix}, \quad \gamma_5 = 
    \begin{pmatrix} \mathds{1}_{2\times 2} & 0\\ 0& -\mathds{1}_{2\times 2} 
    \end{pmatrix}.
\end{equation}

\subsection{Momentum twistor parametrization}
We adopt momentum twistor variables \cite{Hodges:2009hk}, which have proven valuable in studying scattering amplitudes. Momentum twistors automatically satisfy momentum conservation and massless on--shell conditions. Additionally, they hardwire the Gram determinant constraint, allowing us to describe the six--point kinematics in terms of eight unconstrained variables.

For our purposes, it is sufficient to know that for an $n$--particle process, we encode the external kinematics via $n$ momentum twistors
\begin{align}
    Z_i = (\lambda_i, \mu_i),
\end{align}
which are related to invariants built from consecutive sums of momenta via 
\begin{align}
    (p_i+p_{i+1}+\dots + p_{j-1})^2 = \frac{\langle i-1 i j-1 j\rangle}{\langle i-1 i\rangle \langle j-1 j\rangle},
\end{align}
with the momentum twistor four--bracket $\langle ijkl\rangle = \epsilon^{ijkl}Z_i Z_j Z_k Z_l$ and the ordinary spinor brackets $\langle i j\rangle=\langle i j I_\infty\rangle$ which are be expressed in terms of momentum twistors using the infinity bitwistor
\begin{align}
    I_\infty = \begin{pmatrix}0 & 0 \\ 0 & 0 \\ 0& 1 \\ -1 & 0\end{pmatrix}.
\end{align}

In the case of six external momenta, we can use the $\text{SL}(4)$--transformations of the momentum twistors, to pick a particular parametrization in terms of eight independent variables $x_j$,~cf.~\cite{Badger:2013gxa}. To be explicit, in the remainder of this paper we will use

\begin{equation}
\label{eq:t3}
    Z = \begin{pmatrix} 1 & 0 & x_1 & x_1 x_2 & x_1 x_3 & x_1 x_6 \\ 
    0 & 1 & 1 & x_8 & 1 & 1\\
    0 & 0 & 0 & 1 & x_4 & 1\\
    0 & 0 & 1 & 0 & x_5 & x_7
    \end{pmatrix},
\end{equation}
where $\vec{x}=\{x_1,\ldots, x_8\}$ are free variables. The parameterization in \eqref{eq:t3} can be written explicitly as,
\begin{align}
    \label{eq:35}
    s_{12}&=\frac{1}{x_1}, \quad
    s_{23}=\frac{1}{x_1 \left(x_2-x_8\right)},\quad s_{34}=\frac{x_3-x_2 x_4-x_5}{x_1 \left(x_3 x_8-x_2\right)},\notag\\
    s_{45}&=\frac{x_2 \left(x_4(1-x_7)+x_5-1\right)-x_5 \left(x_6+x_8-1\right)-(x_6-x_7)(x_4 x_8-1)+x_3 \left(x_7+x_8-1\right)}{x_1 \left(x_3-x_6\right) \left(x_2-x_8\right)},\notag\\
    s_{56}&=-\frac{x_8 x_5-x_5+x_7-x_4 x_7 x_8}{x_1 \left(x_2-x_3 x_8\right)},\quad s_{61}=-\frac{x_5-x_4 x_7}{x_1 \left(x_3-x_6\right)},\notag\\
    s_{123}&=\frac{x_7+x_8-1}{x_1 \left(x_8-x_2\right)},\quad s_{234}=\frac{x_5}{x_1 \left(x_2-x_3 x_8\right)},\quad
    s_{345}=\frac{x_3-x_5-x_4 x_6+x_4 x_7}{x_1 \left(x_3-x_6\right)}.
\end{align}
All $\eps_{ijkl}$ and $\Delta_6$ are rationalized by this parametrization, e.g.~
\begin{equation}
    \eps_{1234}=-\frac{x_7 \left(x_3+x_4 x_8-1\right)+x_2 \left(x_5+x_4 \left(x_8-x_7\right)-1\right)}{x_1^2 \left(x_2-x_8\right) \left(x_2-x_3 x_8\right)}.
\end{equation}
Note that the parity degree of freedom is reflected in momentum twistor space by the fact that \eqref{eq:35} has two different solutions for any choice of Mandelstam invariants. 

The explicit expression of all kinematic variables in this parameterization is given in the auxiliary file \verb|MomentumTwistorParametrization.m|.

\subsection{Integral families}
\label{ref:subsection_Integral_families}
\begin{figure}[t]
    \centering
    \begin{subfigure}[b]{0.3\textwidth}
    \centering
    \includegraphics[width=\textwidth]{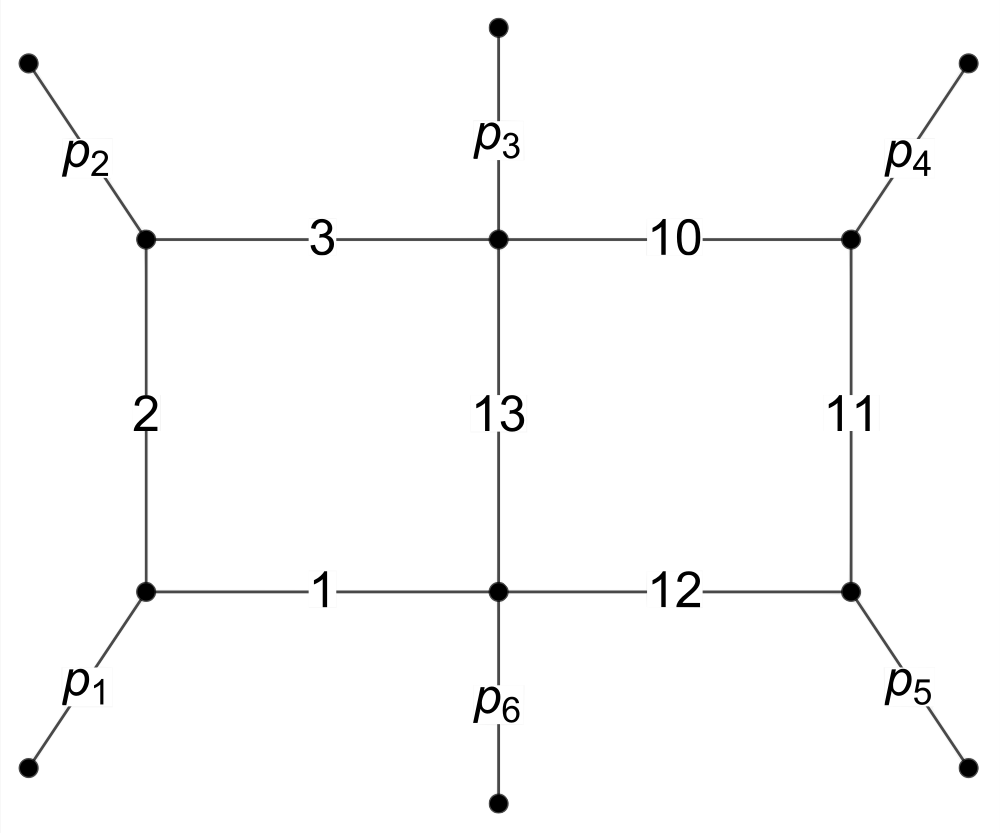}
    \caption{Double-Box (db)}
    \label{fig:double-box}
    \end{subfigure}
    \hfill
    \begin{subfigure}[b]{0.33\textwidth}
    \centering
    \includegraphics[width=\textwidth]{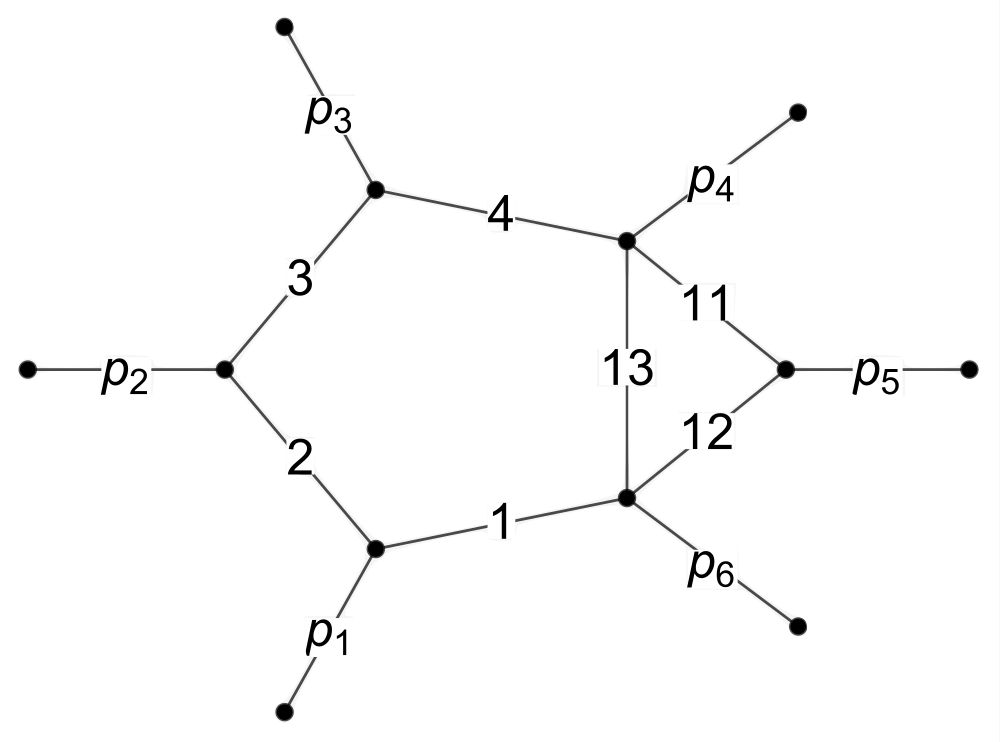}
    \caption{Pentagon-Triangle (pt)}
    \label{fig:pentagon-triangle}
    \end{subfigure}
    \hfill
    \begin{subfigure}[b]{0.25\textwidth}
    \centering
    \includegraphics[width=\textwidth]{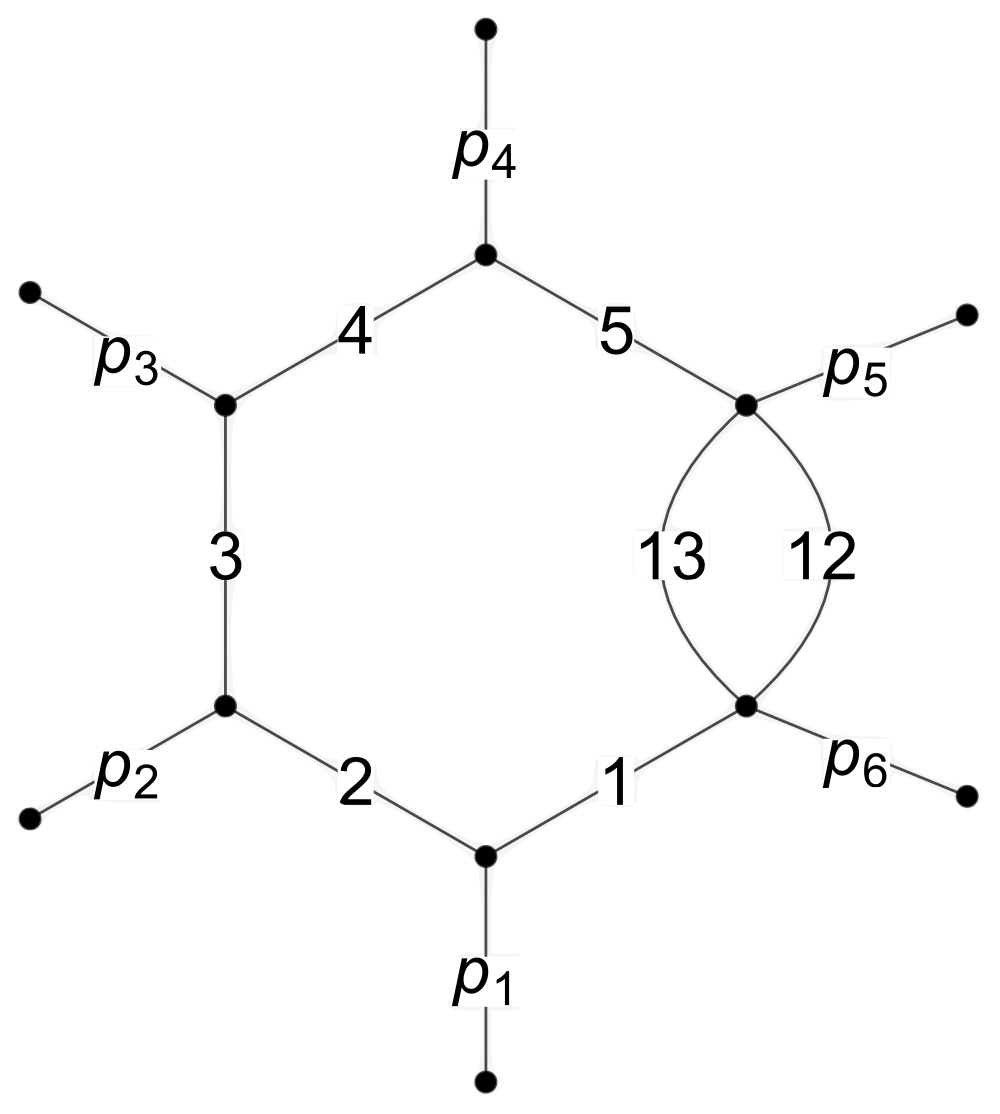}
    \caption{Hexagon-Bubble (hb)}
    \label{fig:hexagon-buble}
    \end{subfigure}
    \caption{Two--loop six--point massless planar diagrams considered in this paper.}
    \label{fig:families}
\end{figure}
All planar two-loop hexagon integrals take the general form
\begin{equation}
    I_{\vec{a}}=\int \frac{d^{4-2\eps}l_1}{i \pi^{2-\eps}}\frac{d^{4-2\eps}l_1}{i \pi^{2-\eps}} \frac{N}{\prod_{j=1}^{13} D_j^{a_j}},
    \label{eq:generalintegral}
\end{equation}
where $a_j\in \mathbb{N}$, $N$ is a numerator that possibly depends on the external and loop momenta, whereas the denominators $D_j$ are chosen as
\begin{align}
D_1&=-l_1^2,\quad D_2=-(l_1-p_1)^2,\quad D_3=-(l_1-p_1-p_2)^2,\quad
 D_4=-(l_1-p_1-p_2-p_3)^2,\nonumber \\
 D_5&=-(l_1-p_1-p_2-p_3-p_4)^2,\quad
 D_6=-(l_1-p_1-p_2-p_3-p_4-p_5)^2,\nonumber\\
D_7&=-l_2^2,\quad D_8=-(l_2+p_1)^2,\quad
D_9=-(l_2+p_1+p_2)^2,\nonumber\\
D_{10}&=-(l_2+p_1+p_2+p_3)^2,\quad 
 D_{11}=-(l_2+p_1+p_2+p_3+p_4)^2,\nonumber \\
  D_{12}&=-(l_2+p_1+p_2+p_3+p_4+p_5)^2,\quad
 D_{13}=-(l_1+l_2)^2.
 \label{ref:propagators}
\end{align}
When $a_i<0$, the corresponding $D_i$ are conventionally called irreducible scalar products, while they are called propagators if $a_i>0$.

Not all integrals of the form \eqref{eq:generalintegral} are independent. Instead, they are related by an infinite set of linear relations with coefficients that are rational functions of the kinematic variables and the dimensional regulator $\eps$. Since these identities arise from the vanishing of integrals whose integrands are given by total derivatives with respect to the loop momenta, they are called integration--by--parts (IBP) identities~\cite{Chetyrkin:1981qh,Laporta:2000dsw}. These identities allow a linear expansion of all integrals of the form \eqref{eq:generalintegral} in terms of a finite set of integrals, which are called master integrals.

In this paper, we study the three families where at most seven of the $a_j$ take positive values. Up to permutations of the external legs, these families can be described by the Feynman diagrams in Figure \ref{fig:families}, which we call top--topologies, and the diagrams which are obtained by pinching some of the internal legs, which we refer to as subtopologies. While the subtopologies correspond to five--point one--mass integrals that have been calculated previously \cite{Abreu:2020jxa}, the calculation of the top--topology integrals is the main goal of this paper. The families are distinguished by which of the $a_j$ are allowed to be non--negative:
\begin{align}
    \text{double--box (db): }&a_1,a_2,a_3,a_{10},a_{11},a_{12},a_{13}\geq0,\\
    \text{pentagon--triangle (pt): }&a_1,a_2,a_3,a_4,a_{11},a_{12},a_{13}\geq0,\\
    \text{hexagon--bubble (hb): }&a_1,a_2,a_3,a_4,a_{5},a_{12},a_{13}\geq0,
\end{align}
For the remaining values of $j$, we have $a_j\leq0$.
Note that the set of propagators \eqref{ref:propagators} is only linearly independent for $D$--dimensional external momenta. If the external momenta are four--dimensional, there are two identities connecting the propagators. This is because e.g.~$p_5\cdot l_1$ and $p_5\cdot l_2$ may be expressed in terms of scalar products between loop momenta and the first four external momenta. We could explicitly use these identities to eliminate two of the above propagators (irreducible scalar products). However, this involves making an arbitrary choice which propagators to remove. This obscures the simple structures of the UT integrals we describe below \ref{sec:PDE}. Instead, we add the corresponding integral identities to the seeds of the IBP system. To be more explicit, we choose to rewrite scalar products between loop momenta and $p_5$ or $p_6$ in terms of the other four momenta according to 
\begin{gather}
    l_j\cdot p_k = \frac{1}{G(1,2,3,4)}\bigg[G\left(
\begin{array}{cccc}
p_k & p_2 & p_3 & p_4 \\
p_1 & p_2 & p_3 & p_4 \\
\end{array}
\right)l_j\cdot p_1+ G\left(
\begin{array}{cccc}
p_1 & p_k & p_3 & p_4 \\
p_1 & p_2 & p_3 & p_4 \\
\end{array}
\right)l_j\cdot p_2
 + \nonumber \\
G\left(
\begin{array}{cccc}
p_1 & p_2 & p_k & p_4 \\
p_1 & p_2 & p_3 & p_4 \\
\end{array}
\right)l_j\cdot p_3
+
G\left(
\begin{array}{cccc}
p_1 & p_2 & p_3 & p_k \\
p_1 & p_2 & p_3 & p_4 \\
\end{array}
\right)l_j\cdot p_4\bigg],
\label{eqn:ISP_reduction}
\end{gather}
for $j=1,2$ and $k=5,6$. We then insert these identities as numerators into the seed integrals and treat them in the same manner as the ordinary IBP identities which are generated using {\sc LiteRed}~\cite{Lee:2012cn}. 

As expected, for all six--point integral families, the number of master integrals with four--dimensional kinematics is lower than the number of master integrals for D--dimensional kinematics. The kinematic identities from relations like \eqref{eqn:ISP_reduction} reduce the number of master integrals. 

\section{Efficiently calculating Feynman integrals}
\label{sec:methodology}
In principle, Feynman integrals can be evaluated numerically using methods such as sector decomposition~\cite{Heinrich:2008si,Heinrich:2023til} or the auxiliary mass flow method (as implemented in AMFlow \cite{Liu:2022chg}). However, for processes with many propagators, these methods become exceedingly computationally expensive and insufficient for phenomenological applications. Additionally, when calculating an integral numerically, one loses the opportunity to learn about the analytic properties of the resulting function, in particular its singularities and branch structure. Therefore, there is strong motivation both from a phenomenological as well as a theoretical point of view to calculate Feynman integrals as analytically as possible.

The method of differential equations \cite{Kotikov:1990kg,Remiddi:1997ny} has proved to be a very powerful tool towards this goal. Of course, being a multivariate function of Mandelstam variables, the value of a Feynman integral is completely determined once we know all of its first derivatives and their values at some arbitrary point in phase space. Since the derivatives of \eqref{eq:generalintegral} with respect to the external variables can be expanded in terms of the same family of integrals, there is a completely algorithmic way to express them in terms of the master integrals of the respective family using IBP identities. However, the swell of intermediate expressions renders this approach unfeasible for state--of--the--art applications. Since this algorithm relies solely on linear algebra involving rational functions of the kinematic variables, it can be accelerated drastically by using finite field reconstructions \cite{vonManteuffel:2014ixa,Peraro:2016wsq,Peraro:2019svx}. In a nutshell, the required calculations are performed for integer--valued kinematics modulo some large prime. Then, once enough information about the resulting rational functions is known, they can be reconstructed in an analytical form. However, in our particular application of eight--scale six--point integrals, the functional reconstruction itself poses a significant bottleneck of this computation. Hence, it is invaluable to provide as much analytic information about the function space as possible, to simplify this final reconstruction step.

Both the reconstruction and the solving of the differential equations is simplified by choosing a basis of master integrals $\vec{I}_\text{fam}$ that consists of pure transcendental functions of uniform transcendental weight \cite{Arkani-Hamed:2010pyv}. Then, the derivatives for the integral family take the form \cite{Henn:2013pwa}
\begin{equation}
    \text{d}\vec{I}_\text{fam}(\eps, \vec{x})= \eps \, \text{d}A_\text{fam}(\vec{x}) \cdot \vec{I}_\text{fam}(\eps,\vec{x}),
    \label{eq:CanonicalPDE}
\end{equation}
Here and in the following we use the subscript $\text{fam}\in\{\text{db},\text{pt},\text{hb}\}$ to distinguish the three families under study, see Figure \ref{fig:families}.
A special feature of eq. (\ref{eq:CanonicalPDE}) is that the dependence on the dimensional regulator $\eps$ is via an overall factor on the right--hand side, so that the matrix $\text{d}A_\text{fam}$ depends on the kinematics only. This matrix is expanded in a basis of dlog forms according to 
\begin{equation}
    \text{d}A_\text{fam}(\vec{x}) = \sum_j c_j^{(\text{fam})} \text{dlog}(\alpha_j(\vec{x}))\,,
    \label{eq:dlogform}
\end{equation}
where the the $\alpha_j$ are algebraic functions of the kinematic variables, whereas the $c_j^{(\text{fam})}$ are block upper triangular matrices whose entries are rational constants. The size of these matrices is determined by the number of master integrals for each of the families. There are 66, 44, and 32 master integrals for the double--box, the pentagon--triangle and the hexagon-bubble family, respectively. The $\alpha_j$ are referred to as letters and together they form what is called the alphabet $\mathds{A}$ of the function space that the integrals evaluate to. The required alphabet for the three families of integrals we are studying is described in section \ref{Sec:Alphabet}.

Let us also note that the question of whether an integral basis can possibly form a canonical basis can be tested efficiently. It is sufficient to perform the finite field evaluation of the differential equation for a small set of kinematic configurations and for different values of $\eps$ to verify the $\eps$--factorized form. Hence this step requires no functional reconstruction. 

Naively, one would reconstruct the rational functions in the derivative~matrices
 \begin{equation}
 \tilde{A}_\text{fam}^j =\frac{\partial A_\text{fam}}{\partial x_j},
 \end{equation}
and integrate those into dlog forms sequentially. In the present case, where some of the leading singularities evaluate to square roots, the square roots have to be extracted from the derivative matrices first. This is straightforward to do, cf.~Sec.~6.3~of~\cite{Peraro:2019svx}. However, the reconstruction can be avoided entirely, if the alphabet $\mathds{A}$ is known a priori \cite{Abreu:2018rcw}. Then, provided a UT basis has been found, the only remaining objects to determine are the $\mathds{Q}$--valued matrices $c_j^{(\text{fam})}$ in equation \eqref{eq:dlogform}. This amounts to a linear fit problem. Since here the final step only consists in a rational reconstruction, instead of a functional reconstruction, this approach requires only a small number of finite field evaluations. 

Ultimately, the combined knowledge of the UT basis and the alphabet, allows for an efficient determination of the differential equations. Calculating the full differential equation in this setup takes around $1.25$ hours, $41$ minutes and $31$ minutes for the double--box, pentagon--triangle and hexagon--bubble, respectively.\footnote{The timings were obtained on \texttt{AMD Ryzen 7 5700G}, with 8 CPU cores and only involve solving the IBPs and the linear fit step.} In contrast, performing a partial functional reconstruction of the dependence on 7 out of the 8 variables on one single entry of the double--box partial differential equation takes roughly 41 hours on a more powerful machine\footnote{We used 4 threads on \texttt{Intel(R) Xenon(R) Platinum 8160 CPU @ $2.10$ GHz} with $800$ GB of memory.}, rendering the whole reconstruction of all 8 partial derivatives for all non-vanishing elements highly inefficient.

\paragraph{Predicting alphabet letters.}

As described above, for an efficient calculation of the multi--scale differential equation, it is invaluable to know the alphabet letters a priori. 

The letters describe the locations of possible singularities of the Feynman integrals, thus they are dictated by the Landau variety \cite{Bjorken:1959fd,Landau:1959fi,Nakanishi:10.1143/PTP.22.128}, see also \cite{Eden:1966dnq}. The letters produced by Landau analysis are rational functions of the kinematic variables, see section~\ref{sec:even letters}. However, it has been observed that in situations where some of the leading singularities of the Feynman integrals under study evaluate to square roots, the alphabet that is required to describe the differential equation in the canonical form also contains algebraic letters of the form 
\begin{equation}
    \alpha_{k} = \frac{P_k-\sqrt{Q_k}}{P_k+\sqrt{Q_k}},
\end{equation}
where both $P_k$ and $Q_k$ are rational functions of the kinematic variables. 

In order to predict the alphabet, we assume that all the even letters, including the occurring square roots~$\sqrt{Q_k}$, needed to fit the differential equation matrices are already known from the literature. As in~\cite{Heller:2019gkq,Zoia:2021zmb, Abreu:2020jxa, FebresCordero:2023gjh}, we assume the factorization property to predict the necessary odd letters. The factorization implies that the product of the denominator and numerator of an odd letter is factorized in terms of even letters,
\begin{equation}
    \left(P_k-\sqrt{Q_k}\right)\left(P_k+\sqrt{Q_k}\right)=c \prod_{i} \alpha_i^{e_i},
    \label{eq:factorization}
\end{equation}
where $c \in \mathbb{Q}$, $e_i \in \mathbb{N}$, $\alpha_{i} \in \mathbb{A}_{\text{even}}$ and $P_k, Q_k \in \mathbb{Z} \left[s_{ij},s_{ijk}\right]$ are homogeneous polynomials of degree $q/2$ and $q$, respectively. This implies that all singularities implied by the vanishing of the odd letters are also supported on solutions of the Landau equations. 

Hence, from the knowledge of the even letters $\alpha_i$ and the occurring square roots~$\sqrt{Q_k}$, the goal is to find all polynomials $P_k$ such that the equation \eqref{eq:factorization} holds,
\begin{equation}
    P_k^2 = Q_k + c \prod_{i} \alpha_i^{e_i}.
    \label{eq:P2}
\end{equation}
Since both $P_k$ and $Q_k$ are homogeneous polynomials, the product of even letters also needs to be a homogeneous polynomial of degree $q$. Therefore, we consider all different products of even letters of a certain degree and check whether \eqref{eq:P2} is a perfect square.

For example, consider the one--loop three--mass triangle integral, cf.~\cite{Zoia:2021zmb}. The even letters appearing in this problem are $\{m_1^2,m_2^2,m_3^2\}$ and the only square root is the leading singularity of the integral $\sqrt{\lambda(m_1^2,m_2^2,m_3^2)}$, where $\lambda(a,b,c)=a^2+b^2+c^2-2ab-2ac - 2bc$. In this case, there are three different solutions to equation \eqref{eq:P2}, namely
\begin{align}
    \lambda(m_1^2,m_2^2,m_3^2) + 4 m_1^2 m_2^2 &= (m_1^2+m_2^2-m_3^2)^2, \notag \\
    \lambda(m_1^2,m_2^2,m_3^2) + 4 m_2^2 m_3^2 &= (-m_1^2+m_2^2+m_3^2)^2, \notag \\
    \lambda(m_1^2,m_2^2,m_3^2) + 4 m_1^2 m_3^2 &= (m_1^2-m_2^2+m_3^2)^2.
\end{align}
Hence, we can construct three different odd letters for this square root. Before we add them to the alphabet, we have to make sure that the letters are multiplicatively independent. In this case, we can check immediately that
\begin{equation}
    1 = \dfrac{m_1^2+m_2^2-m_3^2-\sqrt{\lambda}}{m_1^2+m_2^2-m_3^2+\sqrt{\lambda}}\cdot \dfrac{-m_1^2+m_2^2+m_3^2-\sqrt{\lambda}}{-m_1^2+m_2^2+m_3^2+\sqrt{\lambda}}\cdot \dfrac{m_1^2-m_2^2+m_3^2-\sqrt{\lambda}}{m_1^2-m_2^2+m_3^2+\sqrt{\lambda}}.
\end{equation}Hence, there are only two linearly independent solutions. The full three-mass triangle alphabet reads
\begin{equation}
    \{m_1^2,m_2^2,m_3^2,\sqrt{\lambda}, \dfrac{m_1^2+m_2^2-m_3^2-\sqrt{\lambda}}{m_1^2+m_2^2-m_3^2+\sqrt{\lambda}},\dfrac{-m_1^2+m_2^2+m_3^2-\sqrt{\lambda}}{-m_1^2+m_2^2+m_3^2+\sqrt{\lambda}}\},
\end{equation}
where $\sqrt{\lambda}=\sqrt{\lambda(m_1^2,m_2^2,m_3^2)}$. After relabeling $\{m_1^2 \to s_{12},m_2^2 \to s_{34},m_3^2 \to s_{56}\}$, these two odd letters correspond to the letters $\alpha_{115}$ and $\alpha_{117}$ from section \ref{sec:odd letters}. Let us note that if one embeds the same square root into a bigger alphabet, the number of solutions to equation \eqref{eq:P2} may increase. For example, in the hexagon alphabet presented in \ref{Sec:Alphabet}, the exact same square root also shows up in the letter $\alpha_{119}$.

As the number of even letters grows, the complexity of this problem increases drastically, in particular whenever $\sqrt{Q_k}$ is a product of multiple square roots. For example, the amount of products in terms of the even letters of appropriate polynomial order for the square root $\eps_{1234}\Delta_6$ is roughly $66\cdot10^{15}$. A public version of our implementation to find odd letters efficiently is currently in preparation \cite{Matijasic:2024gkz}.\footnote{We thank Simon Telen for helpful discussions regarding this topic.} 

In the construction of odd letters, we assume that the nine kinematic variables $\vec{s}$ are independent. In this way, we obtain the odd letters for $D$--dimensional external kinematics. Afterwards, we use the momentum twistor parametrization \eqref{eq:35} to satisfy the Gram constraint manifestly and to find relations between these letters for four--dimensional external kinematics. We list these identities in App.~\ref{App:Identities}. 
Thus the odd letters for the four--dimensional case are also obtained.

Ultimately, the proof that we have constructed the full alphabet lies in the success of the linear fit to equation \eqref{eq:dlogform}. If some letters are missing from the ansatz, the corresponding entries of the differential equation can not be determined. However, we find for the three families under study that the letters constructed via \eqref{eq:factorization} are sufficient to describe the entire respective differential equations.

\section{Canonical bases of master integrals}
\label{sec:PDE}


\begin{figure}
    \centering
    \includegraphics[width=0.7\linewidth]{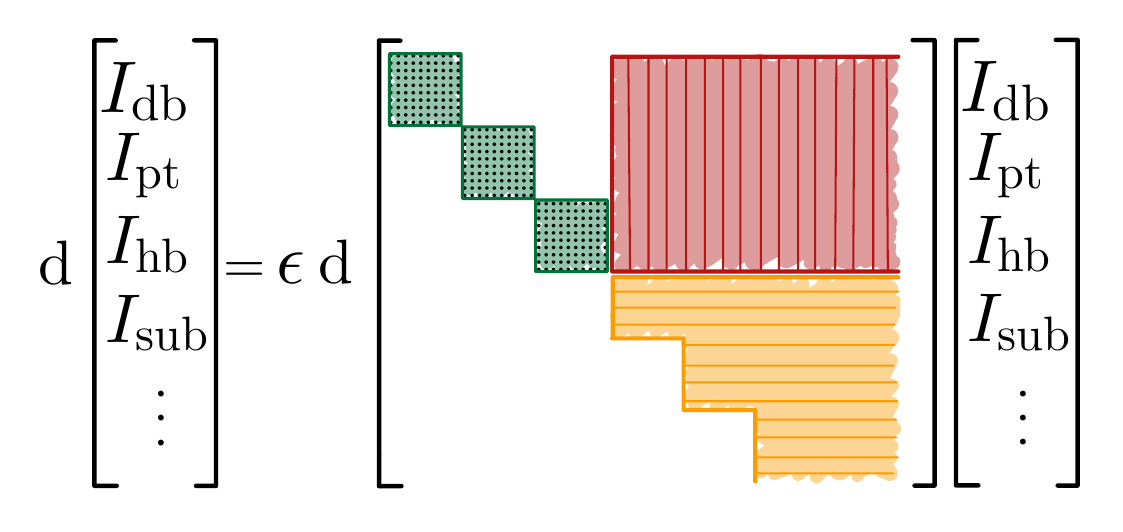}
    \caption{Schematic representation of the structure of the canonical differential equation~\eqref{eq:CanonicalPDE}. The green blocks with a dotted pattern represent the maximal cut blocks computed in Ref.~\cite{Henn:2021cyv}, the yellow blocks with horizontal lines represent the subsector integrals~\cite{Abreu:2020jxa} and the red blocks with vertical lines are the ones computed here.}
    \label{fig:de}
\end{figure}

In Ref.~\cite{Henn:2021cyv} the UT basis for the two--loop hexagon families on their respective maximal cuts was discovered, and the corresponding canonical differential equation on the cut was calculated. This corresponds to the blocks on the diagonal of the differential equation matrix $\text{d}A$ matrix, see Figure \ref{fig:de}. However, the full expression of the UT integrals or the canonical differential equation was unknown beyond the maximal cut. In the remainder of this section, we extend the definitions of the UT integrals beyond their maximal cuts for three of the top--sector families: the double--box, the pentagon--triangle and the hexagon--bubble family (see Figure \ref{fig:families}). In addition to the newly constructed top sector integrals, we use the known subsector integrals from Ref.~\cite{Abreu:2020jxa} to complete the UT basis. The complete canonical differential equations, including the off--diagonal blocks, are extracted using the methods described in section~\ref{sec:methodology}. We provide the differential equations expressed in terms of the alphabet described in section~\ref{Sec:Alphabet} in machine--readable form in the ancillary file \verb|DE_fam.m|.

\subsection{Double--Box family}
Recall that in this paper, we are assuming that the external momenta are four--dimensional. Then, there are seven master integrals in the double--box topsector. Additionally, there are $59$ master integrals in the subsectors of the double--box familiy~\cite{Abreu:2020jxa}. While the integrals proposed in Ref.~\cite{Henn:2021cyv} generate an $\eps$--factorized differential equation on the maximal cut (i.e.~when all subsector integrals are set to zero), this property does not continue to hold beyond the maximal cut. Hence, we need to upgrade the UT basis beyond the maximal cut level. Essentially, the calculations in Ref.~\cite{Henn:2021cyv} are not sensitive to redefinitions of the topsector master integrals which vanish on the maximal cut but might spoil the UT property of the integrals.

In the double--box sector given in Figure (\ref{fig:families}a), the seven double--box UT integrals can be chosen as,
\begin{align}
I_{\text{db},i} & =\eps^4 e^{2 \eps \gamma_{E}}\int \frac{d^{4-2\eps} l_1}{i \pi^{2-\eps}} \frac{d^{4-2\eps} l_2}{i
  \pi^{2-\eps}} \frac{N_i}{D_1 D_2 D_3 D_{10} D_{11} D_{12} D_{13}} , \quad i=1,\ldots,6, \\
  I_{\text{db},7} & =\eps^4 e^{2 \eps \gamma_{E}} \int \frac{d^{4-2\eps} l_1}{i \pi^{2-\eps}} \frac{d^{4-2\eps} l_2}{i
  \pi^{2-\eps}} \frac{N_7}{D_1 D_2 D_3 D_{10} D_{11} D_{12} D_{13}^2} ,
\end{align}
where
\begin{align}
    N_1&= -s_{12} s_{45}s_{156} ,\\
 N_2&= - s_{12} s_{45} (l_1+p_5+p_6)^2 ,\label{ref:double_box_N2}\\
N_3&=\frac{s_{45}}{\epsilon_{5126}}G\left(
\begin{array}{cccc}
l_1& p_1& p_2& p_5+ p_6\\
p_1& p_2& p_5& p_6
\end{array}\right),\label{ref:double_box_N3}\\
N_4&= \frac{s_{12}}{\epsilon_{1543}} G\left(\begin{array}{cccc}
l_2-p_6 & p_5& p_4& p_1+ p_6\\
p_1& p_5& p_4& p_3
\end{array}\right),\\
N_5&=-\frac{1}{4} \frac{\epsilon_{1245}}{G(1,2,5,6)}G\left(
\begin{array}{ccccc}
l_1& p_1& p_2& p_5& p_6\\
l_2& p_1& p_2& p_5& p_6
\end{array}\right),
\end{align}
\begin{align}
N_6&= \frac{1}{8} G\left(
   \begin{array}{ccc}
l_1& p_1& p_2\\
l_2-p_6& p_4& p_5
                   \end{array}\right)+\frac{D_2 D_{11}(s_{123}+s_{126})}{8},\label{ref:double_box_N6} \\
N_7&=-\frac{1}{2 \epsilon }\frac{\hexagonLS}{G(1,2,4,5)}G\left(
\begin{array}{ccccc}
l_1& p_1& p_2& p_4& p_5\\
l_2& p_1& p_2& p_4& p_5
\end{array}\right).
\label{ref:double_box_I7} 
\end{align}

To get the above concise expression of UT integrals, we used two guiding principles: we aim at constructing integrals of definite parity and at writing scalar products in terms of Gram determinants. Even though there is no guarantee that the resulting integrals will be UT, we find empirically that these principles help us find a canonical basis. The UT integrals on the maximal cut that were presented in Ref.~\cite{Henn:2021cyv} are constructed by employing chiral numerators in terms of spinor brackets, hence they transform non-trivially under parity and are only sensible in four--dimensional kinematics. For example, for the double--box top sector, consider a pair of conjugate chiral numerators in terms of spinor helicity formalism,
\begin{align}
   {\mathcal N}_A&= s_{45} \big(\langle 15\rangle
  [52]+\langle 16\rangle [62]\big) l_1 \cdot (\lambda_2\tilde\lambda_1),\\
  {\mathcal N}_B&=s_{45}\big([15]\langle
  52\rangle+[16]\langle 62\rangle\big)  l_1 \cdot (\lambda_1\tilde\lambda_2).
\end{align}
From the six--point four--dimensional kinematics, choose $\{p_1,p_2,p_5,p_6\}$ as a basis and a linear expansion shows,
\begin{align}
  &  (\langle 15\rangle
  [52]+\langle 16\rangle [62]) \lambda_2 \tilde \lambda_1 + ([15]\langle
  52\rangle+[16]\langle 62\rangle) \lambda_1 \tilde \lambda_2\nonumber \\
&=(s_{25}+s_{26}) p_1 + (s_{15}+s_{16}) p_2 -s_{12} p_5 - s_{12} p_6.
\end{align}
Therefore the parity--even combination of two chiral numerators equals,
\begin{gather}
   {\mathcal N}_A+{\mathcal N}_B  =-\frac{1}{2}s_{12} s_{45} (l_1+p_5+p_6)^2 +\frac{1}{2} s_{12} s_{45} s_{156} \nonumber \\+ \bigg(\text{terms proportional to $D_1$,  $\ldots$, $D_7$} \bigg).
\end{gather}

So we use $N_2=s_{12} s_{45} (l_1+p_5+p_6)^2$ in \eqref{ref:double_box_N2} as a numerator to obtain a UT integral. 
The parity--odd combination has the following kinematic relation,
\begin{align}
  \label{eq:3}
 {\mathcal N}_A-{\mathcal N}_B= 
 \frac{-8 s_{45} G\left(
                   \begin{array}{ccccc}
l_1& p_1& p_2& p_5+ p_6\\
p_5& p_1& p_2& p_6
                   \end{array}\right)
}
{\epsilon_{5126}},
\end{align}
so we formulate the expression of $N_3$ in \eqref{ref:double_box_N3} as another numerator to construct a UT integral. The numerator $N_4$  is also a parity--odd combination of chiral numerators. 

The numerator $N_6$ originates from the product of two chiral numerators. The second term in \eqref{ref:double_box_N6} is obtained from transforming the whole differential equation, including the subsectors, to the canonical form. Furthermore, $N_7$ is discovered from the on--cut IBP reduction to the maximally cut UT basis of the double box in \cite{Henn:2021cyv}.

We checked that with the new choice of the basis, the differential equation of the double box sector and its subsectors is canonical.

\subsection{Penta-Triangle family}
In four--dimensional kinematics, there is one master integral in the top sector. An appropriate choice of uniform transcendentality integral is
\begin{align}
    I_\text{pt}=\eps^4 e^{2 \eps \gamma_{E}} \int \frac{d^{4-2\eps}l_1}{i \pi^{2-\eps}}\frac{d^{4-2\eps}l_2}{i \pi^{2-\eps}} \frac{N_{\text{pt}}}{D_1 D_2 D_3 D_4 D_{11}D_{12}D_{13}},
\end{align}
with numerator
\begin{equation}
N_{\text{pt}}= \frac{1}{32} \frac{G\left(
\begin{array}{ccccc}
l_1& p_1& p_2& p_3& p_5\\
l_1& p_1& p_2& p_3& p_5
\end{array}\right)}{\eps_{1235}}.
\end{equation}
Additionally, there are 43 integrals from subtopologies. 

\subsection{Hexagon-Bubble family}
In four--dimensional kinematics, there is one master integral in the top sector. An appropriate choice of uniform transcendentality integral is
\begin{align}
    I_\text{hb}=\eps^3 e^{2 \eps \gamma_{E}} \int \frac{d^{4-2\eps}l_1}{i \pi^{2-\eps}}\frac{d^{4-2\eps}l_2}{i \pi^{2-\eps}} \frac{N_{\text{hb}}}{D_1 D_2 D_3 D_4 D_5 D_{12}D_{13}^2},
\end{align}
with numerator
\begin{equation}
N_{\text{hb}}= \frac{1}{32} \dfrac{G\left(
\begin{array}{ccccc}
l_1& p_1& p_2& p_3& p_4\\
l_1& p_1& p_2& p_3& p_4
\end{array}\right) D_6}{\eps_{1234}}.
\end{equation}
Additionally, there are 31 integrals from subtopologies. 
\section{The alphabet}
\label{Sec:Alphabet}
The analytical structure of the function space of two-loop hexagon integrals is encoded in the hexagon alphabet $\mathbb{A}$. 

The hexagon alphabet consists of letters known from two--loop five--point integrals with one off--shell leg \cite{Abreu:2020jxa} rewritten in terms of six--particle kinematics, letters from the maximal cut \cite{Henn:2021cyv}, letters from the one--loop hexagon integrals \cite{Henn:2022ydo}, and new letters appearing in the off--diagonal blocks of the differential equations.

The alphabet presented in the next sections is closed under the action of dihedral transformations. Hence, it will be sufficient not only to describe the integrals studied in this paper but also all of their dihedral images. The dihedral group is generated by cyclic permutations $T$ defined via
\begin{equation}
    T(p_i)=p_{i+1},\quad i=1,\dots,6,  \quad (p_7 \equiv p_1),
    \label{eq:cyclic_permutations}
\end{equation}
and reflections $\rho$
\begin{equation}
    \rho(p_i)=p_{8-i}, \quad i=1,\dots,6,
\end{equation}
where the indices on the right--hand side are defined modulo 6. 

With permutations included, the two--loop hexagon alphabet has 223 letters which we classify according to parity transformations and list in the following sections. The subset of letters appearing in the differential equations for different families under consideration are given in Table \ref{tab:letters}. 

Note that the differential equations \eqref{eq:CanonicalPDE} are valid for four--dimensional kinematics but we provide the letters in terms of nine scalar products $s_{ij}$. This allows for shorter expressions which are easily converted to four--dimensional kinematics using the map \eqref{eq:35}. While the alphabet we provided in this section is multiplicatively independent as functions of the nine Mandelstam variables, there is a set of identities (App.~\ref{App:Identities}) that hold on the support of the Gram determinant constraint \eqref{eq:GramConstraint}. 

Of course, the well-known nine--letter alphabet for the six--point remainder function in $\mathcal{N}=4$ sYM theory \cite{Dixon:2011ng,Dixon:2011pw} is contained within our alphabet.

\subsection{Parity--even letters}
\label{sec:even letters}
The first part of the alphabet is parity--even letters that are given as scalar products of external momenta.  

There are 48 letters linear in the Mandelstam variables $s_{ij}$:
\begin{align}
    \alpha_{1}&=s_{12}, & \alpha_{i+1}&=T^{i} \alpha_{1}, & i&=1,\ldots,5, \\
    \alpha_{7}&=s_{123}, & \alpha_{i+7}&=T^{i} \alpha_{7}, & i&=1,\ldots,2, \\
    \alpha_{10}&=-s_{12}-s_{23}, & \alpha_{i+10}&=T^{i} \alpha_{10}, &i&=1,\ldots,5, \\
    \alpha_{16}&=s_{12}-s_{123}, & \alpha_{i+16}&=T^{i} \alpha_{16}, &i&=1,\ldots,5, \\
    \alpha_{22}&=s_{12}-s_{345}, & \alpha_{i+22}&=T^{i} \alpha_{22}, &i&=1,\ldots,5, \\
    \alpha_{28}&=-s_{12}-s_{23}+s_{123}, & \alpha_{i+28}&=T^{i} \alpha_{28}, &i&=1,\ldots,5, \\
    \alpha_{34}&=s_{12}-s_{34}-s_{123}, & \alpha_{i+34}&=T^{i} \alpha_{34}, &i&=1,\ldots,5, \\
    \alpha_{40}&=s_{12}-s_{56}+s_{345}, & \alpha_{i+40}&=T^{i} \alpha_{40}, & i&=1,\ldots,5, \\
    \alpha_{46}&=s_{12}+s_{45}-s_{123}-s_{345}, & \alpha_{i+46}&=T^{i} \alpha_{46}, &i&=1,\ldots,2.
\end{align}
The following 42 letters are quadratic in the Mandelstam variables $s_{ij}$:
\begin{align}
    \alpha_{49}&=-s_{12}s_{45}+s_{123}s_{345}, \quad
    \alpha_{i+49}=T^{i} \alpha_{49}, \quad i=1,\ldots,2, \\
    \alpha_{52}&=-s_{12}s_{45}-s_{23}s_{345}+s_{123}s_{345}, \quad
    \alpha_{i+52}=T^{i} \alpha_{52}, \quad i=1,\ldots,5, \\
    \alpha_{58}&=-s_{12}s_{45}-s_{34}s_{123}+s_{123}s_{345}, \quad \alpha_{i+58}=T^{i} \alpha_{58}, \quad i=1,\ldots,5, \\
    \alpha_{64}&=-s_{12}(s_{34}+s_{45})-(s_{34}-s_{56}+s_{123})s_{345}, \quad \alpha_{i+64}=T^{i}
    \alpha_{64}, \quad i=1,\ldots,5, \\
    \alpha_{70}&=(s_{12}+s_{23})s_{45}+s_{123}(s_{61}-s_{23}-s_{345}), \quad \alpha_{i+70}=T^{i} \alpha_{70},\quad  i=1,\ldots,5, \\
    \alpha_{76}&=s_{12}(-s_{34}+s_{345})+(s_{34}-s_{56}-s_{345})s_{345}
, \quad \alpha_{i+76}=T^{i} \alpha_{76}, \quad i=1,\ldots,5, \\
    \alpha_{82}&=(s_{34}-s_{12}+s_{123})(s_{12}-s_{34}+s_{234})-s_{23}s_{56}, \notag\\
    \alpha_{i+82}&=T^{i} \alpha_{82}, \quad i=1,\ldots,5, 
\end{align}
while another six letters are cubic in the scalar products $s_{ij}$:
\begin{align}
    \alpha_{88}&=s_{23}s_{56}(-s_{34}+s_{345})-(s_{61}-s_{234})(s_{12}s_{45}+s_{34}s_{123}-s_{123}s_{345}), \notag \\
    \alpha_{i+88}&=T^{i} \alpha_{88}, \quad i=1,\ldots,5. 
\end{align}
The next five letters are square roots which remain square roots even in the momentum twistor parametrization. They are defined as:
\begin{align}
    \alpha_{94}&=r_1=\sqrt{\lambda(s_{12},s_{34},s_{56})}, \\
    \alpha_{95}&=r_2=\sqrt{\lambda(s_{23},s_{45},s_{61})}, \\
    \alpha_{96}&=r_3=\sqrt{\lambda(s_{12},s_{36},s_{45})}, \notag \\
     \alpha_{i+96}&=r_{i+3}=T^{i} \alpha_{96}, \quad i=1,\ldots,2,
\end{align}
where $\lambda$ denotes the K\"all\'en function
\begin{equation}
    \lambda(a,b,c)=a^2+b^2+c^2-2ab-2ac - 2bc.
\end{equation}
These letters change  sign under the $\sqrt{\lambda} \to -\sqrt{\lambda}$ transformations but  $d \log{(\sqrt{\lambda})}$ is invariant. 

The remaining 16 even letters are pseudo scalars:
\begin{align}
    \alpha_{99}&=\eps_{1234}, & \alpha_{i+99}&=T^{i} \alpha_{99},& i&=1,\ldots,5, \\ 
    \alpha_{105}&=\eps_{1235}, & \alpha_{i+105}&=T^{i} \alpha_{105}, & i&=1,\ldots,5, \\ 
    \alpha_{111}&=\eps_{1245}, & \alpha_{i+111}&=T^{i} \alpha_{111}, & i&=1,\ldots,2, \\ 
    \alpha_{114}&=\hexagonLS.
\end{align}
A pseudo scalar transforms as $\alpha_j \to - \alpha_j$ under parity transformations, but again  $d \log{\alpha_j}$ is invariant. Furthermore, all of the above pseudo scalars are square roots in terms of Mandelstam variables $s_{ij}$ which become rationalized in momentum twistor variables $x_i$. 

To summarize, the parity--even part of the alphabet contains 114 letters closed under the action of the dihedral symmetry group. All of these letters are already known, either from the two--loop six--point maximal cut differential equations or from the two--loop five--point integrals with one off--shell leg. 

Recently, there has been a lot of progress in finding even letters directly from the study of the Landau variety~\cite{Fevola:2023kaw,Fevola:2023fzn,Helmer:2024wax}. We compared the even letters listed in this section with the components of the Landau singular locus of the three families, computed using principal Landau determinants implemented in \verb|PLD.jl|~\cite{Fevola:2023kaw,Fevola:2023fzn}. As already noted in the above references, the current implementation of principle Landau determinants can miss some of the components of the singular locus. For the hexagon--bubble family, all letters are correctly predicted, while for the pentagon--triangle and the double--box family there are 9 and 8 letters respectively that appear in the DE but are not predicted by the package. We list the additional letters needed for the DE in Table~\ref{tab:PLD}.

\begin{table}[t]
\centering
\begin{tabular}{| c |c|c|c|}
\hline
            \multirow{2}{*}{Family} & \# even letters&   \# even letters &  \multirow{2}{*}{additional letters} \\
                  &   in DE    &  from \verb|PLD.jl|     &  \\ \hline
                Hexagon--box (hb) & 48   & 48 (+1) & -- \\ \hline 
                \multirow{2}{*}{Pentagon--triangle (pt)} & \multirow{2}{*}{57} & \multirow{2}{*}{48 (+1)}   & $\alpha_{13}, \alpha_{36},\alpha_{45}, \alpha_{56},\alpha_{58},$ \\
                  &      &      &  $\alpha_{65},\alpha_{73},\alpha_{88},\alpha_{105}$\\ \hline
                \multirow{2}{*}{Double--box (db)} & \multirow{2}{*}{56} & \multirow{2}{*}{48 (+1)}  & $\alpha_{36}, \alpha_{39}, \alpha_{41}, \alpha_{44},$ \\
                  &         &      & $\alpha_{84}, \alpha_{87}, \alpha_{96}, \alpha_{111}$ \\ \hline
\end{tabular}
\caption{Comparison between the even letters predicted by \texttt{PLD.jl} and even letters that show up in the differential equations. The package predicts the same singularities for all three families (and the Gram determinant constraint, represented by the (+1) in the table).}
\label{tab:PLD}
\end{table}

\subsection{Parity--odd letters}
\label{sec:odd letters}
This section lists the parity--odd letters that transform as $d \log(\alpha_i) \to -d\log(\alpha_i)$ under parity transformations.

First, there are 25 letters that change their sign under the sign change of a square root~$r_i$:
\begin{align}
   \alpha_{115}&=\dfrac{s_{12}+s_{34}-s_{56}-r_1}{s_{12}+s_{34}-s_{56}+r_1}, &  \alpha_{116}&=T \alpha_{115}, \\ 
   \alpha_{117}&=\dfrac{-s_{12}+s_{34}+s_{56}-r_1}{-s_{12}+s_{34}+s_{56}+r_1}, &  \alpha_{118}&=T \alpha_{117}, \\ 
   \alpha_{119}&=\dfrac{s_{12}-s_{34}+s_{56}-2s_{123}-r_1}{s_{12}-s_{34}+s_{56}-2s_{123}+r_1}, & \alpha_{i+119}&=T^{i} \alpha_{119}, & i&=1,\ldots,5, \\ 
   \alpha_{125}&=\dfrac{s_{123}+s_{345}-r_3}{s_{123}+s_{345}+r_3}, & \alpha_{i+125}&=T^{i} \alpha_{125}, & i&=1,\ldots,2, \\ 
   \alpha_{128}&=\dfrac{s_{123}-s_{345}-r_3}{s_{123}-s_{345}+r_3}, & \alpha_{i+128}&=T^{i} \alpha_{128}, & i&=1,\ldots,2, \\ 
   \alpha_{131}&=\dfrac{s_{123}+s_{345}-2s_{12}-r_3}{s_{123}+s_{345}-2s_{12}+r_3}, & \alpha_{i+131}&=T^{i} \alpha_{131}, & i&=1,\ldots,2, \\ 
   \alpha_{134}&=\dfrac{s_{123}-s_{345}+2s_{34}-2s_{56}-r_3}{s_{123}-s_{345}+2s_{34}-2s_{56}+r_3}, & \alpha_{i+134}&=T^{i} \alpha_{134}, & i&=1,\ldots,2\\
   \alpha_{137}&=\dfrac{s_{123}-s_{345}-2s_{23}+2s_{61}-r_3}{s_{123}-s_{345}-2s_{23}+2s_{61}+r_3}, & \alpha_{i+137}&=T^{i} \alpha_{137}, & i&=1,\ldots,2
\end{align}
These letters are already known from the two--loop five--particle integrals with one off--shell leg.

In addition, there are 72 letters that are odd with respect to the sign change of the pseudo scalars $\eps_{ijkl}$ and $\hexagonLS$:
\begin{align}
    \alpha_{140}&=\dfrac{s_{12}s_{23}-s_{23}s_{34}+s_{23}s_{56}+s_{34}s_{123}-s_{234}(s_{12}+s_{123})-\eps_{1234}}{s_{12}s_{23}-s_{23}s_{34}+s_{23}s_{56}+s_{34}s_{123}-s_{234}(s_{12}+s_{123})+\eps_{1234}}, \notag \\
    \alpha_{i+140}&=T^{i} \alpha_{140}, \quad i=1,\ldots,5, \\
    \alpha_{146}&=\dfrac{s_{12}(s_{23}-s_{234})-s_{23}(s_{34}+s_{56})+s_{123}(s_{34}+s_{234})-\eps_{1234}}{s_{12}(s_{23}-s_{234})-s_{23}(s_{34}+s_{56})+s_{123}(s_{34}+s_{234})+\eps_{1234}}, \notag \\
    \alpha_{i+146}&=T^{i} \alpha_{146}, \quad i=1,\ldots,5,\\
    \alpha_{152}&=\dfrac{s_{12}(-s_{23}+s_{234})+s_{23}(s_{34}+s_{56})+s_{123}(s_{34}-s_{234})-\eps_{1234}}{s_{12}(-s_{23}+s_{234})+s_{23}(s_{34}+s_{56})+s_{123}(s_{34}-s_{234})+\eps_{1234}}, \notag \\
    \alpha_{i+152}&=T^{i} \alpha_{152}, \quad i=1,\ldots,5, \\
    \alpha_{158}&=\dfrac{s_{12}(s_{23}+s_{234})+s_{23}(-s_{34}+s_{56})+s_{123}(s_{34}-s_{234})-\eps_{1234}}{s_{12}(s_{23}+s_{234})+s_{23}(-s_{34}+s_{56})+s_{123}(s_{34}-s_{234})+\eps_{1234}}, \notag \\
    \alpha_{i+158}&=T^{i} \alpha_{158}, \quad i=1,\ldots,5,\\
    \alpha_{164}&=\dfrac{s_{12}(s_{23}+s_{234})-s_{23}(s_{34}+s_{56}-2s_{234})+s_{123}(s_{34}-s_{234})-\eps_{1234}}{s_{12}(s_{23}+s_{234})-s_{23}(s_{34}+s_{56}-2s_{234})+s_{123}(s_{34}-s_{234})+\eps_{1234}}, \notag \\
    \alpha_{i+164}&=T^{i} \alpha_{164}, \quad i=1,\ldots,5, \\
    \alpha_{170}&=\dfrac{s_{12}(s_{23}-s_{234})+s_{23}(-s_{34}+s_{56}-2s_{123})+s_{123}(-s_{34}+s_{234})-\eps_{1234}}{s_{12}(s_{23}-s_{234})+s_{23}(-s_{34}+s_{56}-2s_{123})+s_{123}(-s_{34}+s_{234})+\eps_{1234}}, \notag \\
    \alpha_{i+170}&=T^{i} \alpha_{170}, \quad i=1,\ldots,5, \\
    \alpha_{176}&=\dfrac{s_{12}(s_{23}+2s_{56}-2s_{123}-s_{234})+s_{23}(-s_{34}+s_{56})+s_{123}(s_{34}-s_{234})-\eps_{1234}}{s_{12}(s_{23}+2s_{56}-2s_{123}-s_{234})+s_{23}(-s_{34}+s_{56})+s_{123}(s_{34}-s_{234})+\eps_{1234}}, \notag \\
    \alpha_{i+176}&=T^{i} \alpha_{176}, \quad i=1,\ldots,5, \\
    \alpha_{182}&=\dfrac{2s_{12}^2+s_{12}(s_{23}-2s_{34}-2s_{123}+s_{234})+s_{23}(-s_{34}+s_{56})+s_{123}(s_{34}-s_{234})-\eps_{1234}}{2s_{12}^2+s_{12}(s_{23}-2s_{34}-2s_{123}+s_{234})+s_{23}(-s_{34}+s_{56})+s_{123}(s_{34}-s_{234})+\eps_{1234}}, \notag \\
    \alpha_{i+182}&=T^{i} \alpha_{182}, \quad i=1,\ldots,5.
\end{align}
These 48 letters are known from the one--loop hexagon integral and the two--loop five--point integrals with one massive leg.

The following six letters appear for the first time in the off--diagonal block of the  differential equations for pentagon--triangle integral family,
\begin{align}
    \alpha_{188}&=\dfrac{-s_{12}(s_{45}+s_{61}-s_{234})+s_{23}(s_{34}+s_{56}-s_{345})+s_{123}(-s_{34}+s_{61}-s_{234}+s_{345})-\eps_{1235}}{-s_{12}(s_{45}+s_{61}-s_{234})+s_{23}(s_{34}+s_{56}-s_{345})+s_{123}(-s_{34}+s_{61}-s_{234}+s_{345})+\eps_{1235}}, \notag \\
    \alpha_{i+188}&=T^{i} \alpha_{188}, \quad i=1,\ldots,5.
\end{align}
Furthermore, the letters $\{\alpha_{194}, \ldots, \alpha_{205}\}$ appear for the first time in the off--diagonal block of the double--box differential equation. These are defined as
\begin{gather}
    \alpha_{194}=\dfrac{P_1-\eps_{1245}}{P_1+\eps_{1245}},  \quad
    \alpha_{i+194}=T^{i} \alpha_{194}, \quad i=1,2, \\
    \alpha_{197}=\dfrac{P_2-\eps_{1245}}{P_2+\eps_{1245}},  \quad
    \alpha_{i+197}=T^{i} \alpha_{197}, \quad i=1,2, \\
    \alpha_{200}=\dfrac{P_3-\eps_{1245}}{P_3+\eps_{1245}}, \quad
    \alpha_{i+200}=T^{i} \alpha_{200}, \quad i=1,2, \\
    \alpha_{203}=\dfrac{P_4-\eps_{1245}}{P_4+\eps_{1245}}, \quad
    \alpha_{i+203}=T^{i} \alpha_{203}, \quad i=1,2,
\end{gather}
where $P_1$, $P_2$, $P_3$ and $P_4$ are polynomials defined as
\begin{align}
    P_{1}&=s_{12}s_{45}+s_{23}(-s_{34}+s_{56}+s_{345})+s_{61}(s_{34}-s_{56})+s_{123}(s_{34}+s_{61}-s_{234}) \notag \\
    &\qquad +s_{345}(s_{56}-s_{123}-s_{234}),\\
    P_{2}&=s_{12}s_{45} - s_{34}s_{61} +s_{56}s_{61} - s_{34}s_{123}+s_{123}(-s_{61}+ s_{234} + s_{345}) \notag \\
   &\qquad + s_{23} (s_{34} -s_{56} - s_{345})-s_{56}s_{345} + s_{234}s_{345},\\
    P_{3}&=s_{12}s_{45}+s_{123}(-s_{34}+s_{123})+2s_{23}^2+s_{23}\left(s_{34}+s_{56}-2(s_{61}+s_{123}+s_{234})+s_{345}\right) \notag\\
    &\qquad +s_{61}(-s_{34}-s_{56}+s_{123}+2s_{234})+s_{345}(s_{56}-s_{123}-s_{234}),\\
    P_{4}&= s_{12}s_{45} + s_{23}(s_{34}-s_{56}-s_{345}) + 2 s_{34}^2 +s_{34} (-2 s_{56} +s_{61} +s_{123} -2 (s_{234}+s_{345})) \notag \\
    &\qquad-s_{56}s_{61} + s_{61}s_{123} + 2 s_{56}s_{234} - s_{123}s_{234} + s_{345} (s_{56} - s_{123} +s_{234}).
\end{align}
The remaining three odd letters are known from the one--loop hexagon integral,
\begin{align}
    \alpha_{206}&=\dfrac{-s_{12}s_{45}s_{234}+s_{34}s_{61}s_{123}+s_{345}(-s_{23}s_{56}+s_{123}s_{234})-\hexagonLS}{-s_{12}s_{45}s_{234}+s_{34}s_{61}s_{123}+s_{345}(-s_{23}s_{56}+s_{123}s_{234})+\hexagonLS}, \notag \\
    \alpha_{i+206}&=T^{i} \alpha_{206}, \quad i=1,\ldots,2.
\end{align}

\subsection{Letters with mixed parity transformations}

The last 15 letters in the hexagon alphabet transform non--trivially under the parity transformations. They are even under the simultaneous change of the sign of a square root $r_i$ and a pseudo scalar $\eps_{ijkl}$, but odd under the sign change of just one of them.

The letters are defined as:
\begin{align}
    \alpha_{209}&=\dfrac{P_5-r_1\eps_{1234}}{P_5+r_1\eps_{1234}}, \notag \\
    \alpha_{i+209}&=T^{i} \alpha_{209}, \quad i=1,\ldots,5, \\
    \alpha_{215}&=\dfrac{P_6-r_4\eps_{1234}}{P_6+r_4\eps_{1234}}, \notag \\
    \alpha_{i+215}&=T^{i} \alpha_{215}, \quad i=1,\ldots,5, \\
    \alpha_{221}&=\dfrac{P_7-r_3\eps_{1245}}{P_7+r_3\eps_{1245}}, \notag \\
    \alpha_{i+221}&=T^{i} \alpha_{221}, \quad i=1,\ldots,2,
\end{align}
where $P_5$, $P_6$ and $P_7$ are the following polynomials:
\begin{align}
    P_5&=s_{12}^2(s_{23}-s_{234})+s_{12} \left( -2s_{23} (s_{34} +s_{56}) +s_{34}(-2s_{56}+s_{123}) +s_{234}(s_{34}+s_{56}+s_{123})\right) \notag \\
    &\quad +(s_{34}-s_{56}) \left(s_{23}(s_{34}-s_{56}) +s_{123}(-s_{34}+s_{234}) \right),\\
    P_6&=s_{12} \left(s_{23}(2s_{56}-s_{123}+s_{234}) - s_{234}(s_{123}+s_{234}) \right)  -s_{123}(s_{34}-s_{234})(s_{123}+s_{234})  \notag \\
    &\quad+ 2s_{23}^2s_{56}+ 2s_{23} \left(s_{34}(2s_{56} +s_{123}-s_{234}) -s_{56}(s_{123}+s_{234})-2s_{123}s_{234} \right),\\
    P_7&=s_{12}s_{45} \left(2s_{23}+2s_{34}+2s_{56}+2s_{61}-s_{123} -4s_{234} -s_{345} \right) + (s_{123}+s_{345})\cdot \notag \\
   & \left(s_{23}(s_{34}-s_{56}-s_{345})-s_{61}(s_{34}-s_{56})-s_{123}(s_{34}+s_{61}-s_{234}) -s_{345}(s_{56}-s_{123}-s_{234})  \right).
\end{align}
The first 12 letters are already known from the literature, while the letters $\{\alpha_{221}, \ldots, \alpha_{223} \}$ are new. These new letters appear in the off--diagonal blocks of the double--box differential equations. 

In total, the alphabet identified here has 223 letters, 114 of which are even, 94 odd and 15 with mixed parity behavior. To express the DEs of the particular orientations of integral families shown in Figure~\ref{fig:families}, 101 letters are needed for the double--box family, 95 for the pentagon--triangle family and 100 for the hexagon--bubble family. 
The specific letters appearing in the differential equations for each of the families are listed in Table \ref{tab:letters}.

\section{Solving the differential equations}
In this section, we describe how to solve the canonical differential equations in eq.~\eqref{eq:CanonicalPDE}. We provide boundary constants for all integrals in the three families up to transcendental weight four, fix the full functional dependence in the Euclidean region in terms of classical polylogarithms up to weight two and describe how to define one--fold integral representations for all functions up to weight four, which can be used to provide fast numerical evaluations of the integrals. 

\subsection{Boundary constants}
\label{sec:boundary}
\begin{figure}[t]
    \centering
    \includegraphics[width=0.75\linewidth]{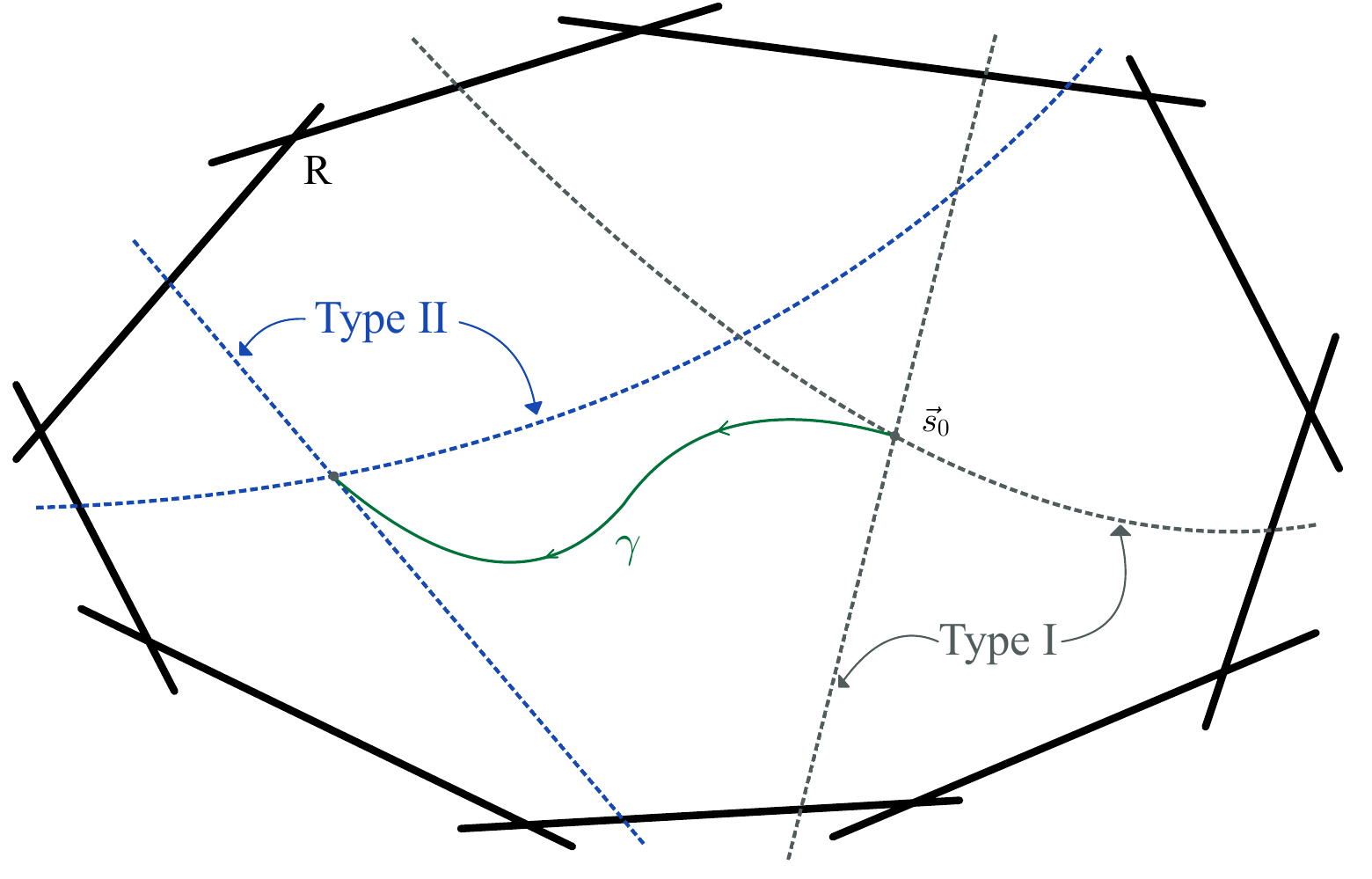}
    \caption{Schematic representation of the Euclidean region R which is bounded by the inequalities $s_{ij}<0$ and $s_{ijk}<0$ (solid black lines). The dashed lines represent different hypersurfaces where some alphabet letters $\alpha_j$ vanish.}
    \label{fig:Euclidean region}
\end{figure}

To uniquely fix a solution of the first--order canonical DE, we have to provide boundary information for all of the integrals at a single point. 
In this paper, we fix the boundary constants analytically by imposing regularity of our basis of integrals throughout the Euclidean region~\cite{Henn:2013nsa,Chicherin:2018mue,Henn:2020lye}. We record the values of our integral bases at the symmetric reference point
\begin{equation}
    \vec{s}_0= \{-1,\dots,-1\}.
\end{equation}
While the basis should be regular all through the Euclidean region, defined by
\begin{align}
    s_i < 0,
\end{align}
this is not true for the differential equation matrices. Instead, the Euclidean region is crossed by the hypersurfaces defined by the vanishing of any alphabet letter
\begin{equation}
    \alpha_j = 0.
\end{equation}
Schematically, the situation is depicted in Figure \ref{fig:Euclidean region}, where solid black lines denote the boundaries of the Euclidean region, while the dashed lines depict the spurious hypersurfaces defined by the respective vanishing of one of the alphabet letters. Imposing the absence of singularities on these spurious surfaces puts constraints on the boundary constants. The constraints arising from the hypersurfaces that contain the boundary point $\vec{s}_0$ are labeled Type--I in Figure \ref{fig:Euclidean region}. The constraints from all other spurious hypersurfaces, which we call Type--II constraints, must first be transported to the boundary point using the differential equation along some path $\gamma$.

 We find that solving the DEs on the one--parameter curve defined by
\begin{align}
    s_{12}&=s_{34}=s_{56}=-1,\notag\\
    s_{23}&=s_{45}=s_{61}=-(x-1)^2,\notag\\
    s_{123}&= s_{234}= s_{345}=x-1, 
    \label{eq:curve}
\end{align}
and requiring regularity at the points $x=0,\rho,\bar{\rho}$ with $\rho = \frac{1}{2}(1-i \sqrt{3})$
is sufficient to fix all boundary constants up to weight four up to an overall rescaling. The latter is fixed once we substitute the known expansion for our UT choice for the sunrise integrals
\begin{align}
   I_\text{sunrise}(\vec{s}_0) &= -e^{2 \gamma_E \eps} \frac{\Gamma(1-\eps)^3 \Gamma(1+2\eps)}{2\Gamma(1-3\eps)}\notag\\
   &=-\frac{1}{2}+ \frac{\pi^2}{12}\eps^2 + \frac{16}{3}\zeta_3\eps^3 + \frac{19 \pi^4}{240}\eps^4+ \mathcal{O}(\eps^5).
\end{align}
On the curve defined by equation \eqref{eq:curve}, the alphabet of all three families under study simplifies massively, such that all remaining non--constant logarithms have arguments from the set
\begin{align}
    \mathbb{A}_\text{line}=\{x, x-1, 1-x+x^2\}.
\end{align}
Since the last letter can be factorized over the complex numbers as 
\begin{equation}
    1-x+x^2 = \left[x-\frac{1}{2}\left(1-i\sqrt{3}\right)\right]\left[x+\frac{1}{2}\left(1-i \sqrt{3}\right)\right]
\end{equation}
we can express the general solution to the DEs on this curve in terms of Goncharov $G$--functions \cite{Goncharov:1998kja} with entries in the set $\{0,1,\rho,\bar{\rho}\}$. Hence, the boundary constants can be expressed in terms of linear combinations of polylogarithms at the sixth roots of unity \cite{Henn:2015sem}. 

Using the function \verb|DecomposeToLyndonWords| from {\sc PolyLogTools} \cite{Duhr:2019tlz} it is straightforward to isolate the log--divergent contributions in the limits $x\rightarrow0,\rho,\bar{\rho}$. Imposing their vanishing order--by--order in $\epsilon$ fixes the boundary constants relative to each other. Here we provide the boundary values for the new top--sector integrals,
\begin{align}
    I_{\text{db},1}(\vec{s}_0) &= 1 + \frac{\pi^2}{6}\eps^2 + \frac{38}{3}\zeta_3\eps^3+\left(\frac{49\pi^4}{216}+ \frac{32}{3}\text{ Im}\left[\text{Li}_2(\rho)\right]^2\right)\eps^4,\notag\\
    I_{\text{db},2}(\vec{s}_0) &= 1 + \frac{\pi^2}{6}\eps^2 + \frac{34}{3}\zeta_3\eps^3+\left(\frac{71\pi^4}{360}+ 20\text{ Im}\left[\text{Li}_2(\rho)\right]^2\right)\eps^4,\notag\\
    I_{\text{db},3}(\vec{s}_0) &= I_{\text{db},4}(\vec{s}_0) = I_{\text{db},5}(\vec{s}_0) = 0, \notag\\
    I_{\text{db},6}(\vec{s}_0) &= -\left(\frac{\pi^4}{540}+\frac{4}{3}\text{ Im}\left[\text{Li}_2(\rho)\right]^2\right)\eps^4,\notag\\
    I_{\text{db},7}(\vec{s}_0) &= I_{\text{pt}}(\vec{s}_0) = I_{\text{hb}}(\vec{s}_0) = 0. 
\end{align}
 We give machine--readable analytic expressions for the boundary values of all UT integrals in terms of classical polylogarithms in the ancillary file \verb|BoundaryValues.m|. We verified these analytic boundary values by comparing them to the numerical results from {\sc AMFlow}~\cite{Liu:2022chg} and found perfect agreement at 100--digit precision.
\subsection{Analytic properties of the hexagon function space}
The solutions to the canonical differential equations can be found as a series expansion in the dimensional regulator $\eps$. By construction, the expansion starts at $\mathcal{O}(\eps^0)$,
\begin{equation}
    \vec{I}_{\text{fam}}(\vec{s},\eps)=\sum_{k=0}^{\infty}\eps^{(k)} \vec{I}^{\,(k)}_{\text{fam}}(\vec{s}\,),
\end{equation}
where $\vec{I}_{\text{fam}}$ denotes the basis integrals for one of the three families. The weight $k$ solution is then given as $k$--fold iterated integral
\begin{equation}
    \vec{I}_{\text{fam}}^{\,(k)}(\vec{s}\,)=\vec{I}_{\text{fam}}^{(k)}(\vec{s}_0) + \int_{\gamma} (dA(\vec{s}\,'))\vec{I}_{\text{fam}}^{\,(k-1)}(\vec{s}\,'),
    \label{eq:iterated int}
\end{equation}
where $\vec{I}_{\text{fam}}^{\,(k)}(\vec{s}_0)$ are weight $k$ boundary values described in the previous section and $\gamma$ is a path connecting the boundary point $\vec{s}_0$ and some other point $\vec{s}$. 

The integral in equation \eqref{eq:iterated int} can be written as
\begin{equation}
    \vec{I}_{\text{fam}}^{\,(k)} (\vec{s}\,) = \sum_{k'=0}^k\; \sum_{i_1,\ldots,i_{k'}=1}^{223} a^{(a_1)}\ldots a^{(i_{k'})} \vec{I}^{\,(k-k')}_{\text{fam}} (\vec{s}_0) \left[ \alpha_{i_1},\ldots,\alpha_{i_{k'}}\right]_{\vec{s}_0} (\vec{s}\,),
    \label{eq:Chen iterated int}
\end{equation}
where we use the recursive definition of Chen iterated integrals~\cite{Chen:1977oja} of weight $k$
\begin{equation}
    \left[ \alpha_{i_1},\ldots,\alpha_{i_{k}}\right]_{\vec{s}_0} (\vec{s}\,) = \int_{\gamma} \text{d}\log \alpha_{i_k} (\vec{s}\,') \left[ \alpha_{i_1},\ldots,\alpha_{i_{k-1}}\right]_{\vec{s}_0} (\vec{s}\,')
\end{equation}
with $[\,]_{\vec{s}_0}=1$. The differential equations ensure that only homotopy invariant linear combinations of iterated integrals appear in the solution \eqref{eq:Chen iterated int}, while a single term is not homotopy invariant. 
 \begin{table}[t]
    \centering
    \begin{tabular}{| c | c | c | c |}
        \hline
         & Double--box & Pentagon--triangle & Hexagon--bubble \\ [0.5ex]
        \hline
        \multirow{3}{4em}{Weight 1} & & &\\ 
        	& $\alpha_{1}, \ldots, \alpha_{9}$ & $\alpha_{1} , \ldots, \alpha_{9}$ & $\alpha_{1} , \ldots, \alpha_{9}$ \\
        	& & & \\ [0.5ex]
        \hline
        \multirow{3}{4em}{Weight 2} & $\alpha_{16} , \ldots,\alpha_{27}$, & $\alpha_{16} , \ldots, \alpha_{27}$, & $\alpha_{16} , \ldots, \alpha_{27}$,\\
        	& $\alpha_{49} , \ldots, \alpha_{51}$, & $\alpha_{49}, \ldots,\alpha_{51}$, & $\alpha_{49}, \ldots, \alpha_{51}$,\\
       		& $\alpha_{115} , \ldots, \alpha_{118}$ & $\alpha_{115}, \ldots, \alpha_{118}$ & $\alpha_{115} , \ldots, \alpha_{118}$\\ [0.5ex]
       	\hline
       	\multirow{9}{4em}{Weight 3} & $\alpha_{28} , \ldots, \alpha_{33}$,& $\alpha_{28} , \ldots,\alpha_{33}$ &  $\alpha_{28} , \ldots, \alpha_{33}$,\\
       	& $\alpha_{46} , \ldots, \alpha_{48}$, & $\alpha_{46} , \ldots, \alpha_{48}$, &  $\alpha_{46} , \ldots, \alpha_{48}$,\\
       		& $\alpha_{76} , \ldots,\alpha_{81}$, $\alpha_{94},\, \alpha_{95}$, &$\alpha_{56},\, \alpha_{58}$, $\alpha_{76}, \ldots, \alpha_{81}$, & $\alpha_{76},\ldots, \alpha_{81}$,\,  \\
       		&$\alpha_{119} , \ldots, \alpha_{124}$, &$\alpha_{94},\, \alpha_{95}$, $\alpha_{119}, \ldots, \alpha_{124},$ & $\alpha_{94}$, $\alpha_{95}$,  \\
         & $\alpha_{140}, \ldots, \alpha_{148},$ &$\alpha_{140}, \ldots,\alpha_{148}$,&$\alpha_{119}, \ldots, \alpha_{124},$ \\
       		&$\alpha_{152},\ldots,\alpha_{154}$, & $\alpha_{152},\ldots,\alpha_{154},$ & $\alpha_{140}, \ldots, \alpha_{157}$,\\
       		&  $\alpha_{164},\ldots \alpha_{169},$ & $\alpha_{164}, \ldots,\alpha_{169}$,&  $\alpha_{164}, \ldots,\alpha_{181}$,\\
         &  $\alpha_{176} , \ldots,\alpha_{178}$, &  $\alpha_{176},\ldots,\alpha_{178}$,&  $ \alpha_{209} , \ldots,\alpha_{214}$ \\
       		&  $\alpha_{209} , \ldots,\alpha_{212}$ &   $\alpha_{206}, \ldots,\alpha_{212}$ &  \\ [0.5ex]
       	\hline
       	\multirow{3}{4em}{Weight 4} & $\alpha_{36},\, \alpha_{39}$, $\alpha_{41}, \,\alpha_{44}$, & $\alpha_{13},\, \alpha_{36}$, $\alpha_{45},\, \alpha_{65}$,& $\alpha_{99}, \ldots,\alpha_{104}$, \\
       		& $\alpha_{99}, \ldots, \alpha_{104}$,\,$\alpha_{114}$, & $\alpha_{73},\, \alpha_{88},\, \alpha_{99}, \ldots,\alpha_{104},$ & $\alpha_{114}$ \\
       		& $\alpha_{125},\, \alpha_{128},\, \alpha_{131},\, \alpha_{214}$ & $\alpha_{114},\,\alpha_{155},\, \alpha_{160},\, \alpha_{188}$ & \\ [0.5ex]
       	\hline
       	\multirow{3}{4em}{Weight 5} & $\alpha_{84},\, \alpha_{87},\, \alpha_{96}$ & & \\
       		& $\alpha_{134},\, \alpha_{137},\, \alpha_{184},$ & $\alpha_{105}$ & -- \\
       		&  $\alpha_{187},\, \alpha_{217},\, \alpha_{220}$ & & \\ [0.5ex]
       	\hline
        \multirow{1}{4em}{Total}& 101 & 95 & 100 \\
       	\hline
    	\end{tabular}
    \caption{List of the alphabet letters appearing in the differential equations for each of the families considered. The letters are classified according to the weight at which they first appear in the symbol. Note that all of the letters from the weight $k-1$ of the symbol also appear at the symbol of weight $k$.}
    \label{tab:letters}
 \end{table}

One useful tool to study the properties of polylogarithmic functions is the symbol. The symbol map $\mathcal{S}$ can be defined by its action on the Chen iterated integrals~\cite{Goncharov:2010jf}
\begin{equation}
    \mathcal{S} \left(\left[ \alpha_{1},\ldots,\alpha_{k}\right]_{\vec{s}_0} (\vec{s}\,) \right) = \alpha_1 \otimes \ldots \otimes \alpha_k,
\end{equation}
which maps the $k$--fold iterated integral into the $k$--fold tensor product. Using the symbol map, we classify the alphabet letters according to the weight at which they appear in the symbol for the first time for each family. We record this information in Table \ref{tab:letters} and use it to predict the function space for the basis integrals at weight one and weight two described in the following section. 
 
Additionally, we confirm the validity of the extended Steinmann relations \cite{Caron-Huot:2019bsq} for all integrals in the basis of all three families considered. These relations state that double discontinuities in partially overlapping channels vanish~\cite{Steinmann:1960a, Steinmann:1960b} and require that the three--point kinematic variables $s_{123}, s_{234}$ and $s_{345}$ never appear next to each other in the symbol. The differential equation matrices $A_{(\text{fam})}$ assure that the Steinmann relations hold at any order in $\eps$ at any depth into the symbol \cite{Chicherin:2020umh}
\begin{equation}
    c_j^{(\text{fam})} \cdot c_k^{(\text{fam})}=0, \quad j \ne k, \quad j,k=7,8,9.
    \label{eq:adjacency}
\end{equation}
In practice, these adjacency conditions imply that any two letters $\alpha_j$ and $\alpha_k$ for which this identity holds will never appear as consecutive letters in the symbol. 

Constraints like these are very important for bootstrap approaches to amplitudes or other observables. If the alphabet for an object is known, one can make an ansatz in terms of all possible functions built from this alphabet and determine the respective coefficients from physical properties of the result, see e.g.~\cite{Dixon:2011pw,Dixon:2011nj,Caron-Huot:2020bkp}. However, for large alphabets and with growing weight, the size of the required function basis grows rapidly. Then, a priori knowledge about the structure of the function space, like the above adjacency conditions, cuts down the size of the required ansatz.
Motivated by this application, we also consider the adjacency constraints for all other pairs of letters in our alphabet. Considering the overlap of all three families, we find $1658$ forbidden pairs of letters. Of course, this number of forbidden pairs is not preserved under basis changes for the function alphabet. Additionally, it is possible that some of these forbidden pairs are accidental at the two--loop level and will not continue to hold at higher loops.

\subsection{One--fold integral representation for numerical evaluations}
Even though, the differential equations can be solved up to any desired order in dimensional regulator $\eps$ using iterated integrals, for physical scattering amplitudes computations, one typically wishes to evaluate them up to the finite part. It is conjectured that an $L$--loop Feynman integral in $D$ dimensions can at most involve transcendental functions of weight $\lfloor \frac{LD}{2}\rfloor$ \cite{Hannesdottir:2021kpd}. Therefore, we are interested in the solution to the canonical differential equation~\eqref{eq:CanonicalPDE} up to weight four in the dimensional regulator $\eps$. 

At weights one and two, we can explicitly integrate the iterated integrals \eqref{eq:iterated int} into special functions such that they are well--defined in the whole Euclidean region. Moreover, the only integrals that contribute at these weights are the ones coming from the subsectors, thus their function space is already known~\cite{Gehrmann:2018yef,Chicherin:2020oor,Chicherin:2021dyp}. At weight one, the only function that can appear is the logarithm
\begin{gather}
    f_i^{(1)}= \log(-s_{i\, i+1}), \quad i=1,\ldots,6,\\
    f_{i+6}^{(1)}= \log(-s_{i\, i+1 \, i+2}), \quad i=1,\ldots,3.
\end{gather}

The function basis at weight two is given by genuine weight two functions and products of weight one functions. Genuine weight two functions that appear are:
\begin{align}
    f_1^{(2)} &= \text{Li}_2 \left(1 - \dfrac{s_{12}}{s_{123}}\right), \quad f_{i+1}^{(2)}=T^i f_1^{(2)}, \quad i=1,\ldots,5, \notag \\
    f_7^{(2)} &= \text{Li}_2 \left(1 - \dfrac{s_{23}}{s_{123}}\right), \quad f_{i+7}^{(2)}=T^i f_7^{(2)}, \quad i=1,\ldots,5, \notag \\
    f_{13}^{(2)} &= \text{Li}_2 \left(1 - \dfrac{s_{12} s_{45}}{s_{123} s_{345}}\right), \quad f_{i+13}^{(2)}=T^i f_{13}^{(2)}, \quad i=1,\ldots,2, \notag \\
    f_{16}^{(2)}&=\text{Tri}(s_{12},s_{34},s_{56}), \quad f_{17}^{(2)}=\text{Tri}(s_{23},s_{45},s_{61}),
\end{align}
where the $\text{Tri}(a,b,c)$ function is the Bloch--Wigner dilogarithm
\begin{align}
\text{Tri}(a,b,c) = \left.\text{BW}(z,\bar{z})\right|_{z \bar z = a/b,(1-z)(1-\bar{z})=c/b}
\end{align}
defined via
\begin{align}
    \text{BW}(z,\bar{z}) = 2\text{Li}_2(z)-2\text{Li}_2(\bar{z}) + \log(z\bar{z})\left[\log{(1-z)} - \log{(1-\bar{z})}\right].
\end{align} 

This basis of special functions is sufficient to express the solution up to weight two for any master integral in our basis for the three families we are considering.

We still need to get weight three and four parts of the solution. We could try to find a similar basis of functions at higher weight consisting of $\text{Li}_3, \text{Li}_4$ and $\text{Li}_{2,2}$ functions.
Finding such a basis of functions is not easy, since is not generally understood which arguments to use in order for them to form a basis and to be well--defined in a desired kinematic region. Moreover, at higher weights, there is a proliferation of terms which may lead to a slowdown of numerical evaluations.
Following Ref.~\cite{Caron-Huot:2014lda}, we employ a hybrid approach. Since we know the boundary value at the point $\vec{s}_0$ up to weight four, we can use that information to set up a one--fold integration over the weight--two functions along a straight line to some other point within the Euclidean region. This representation is well--suited for fast numerical evaluations with high precision.

We start by setting up a straight line between our starting point $\vec{s}_0$ and an endpoint. When considering such a path, we need to make sure that the path satisfies the Gram constraint \eqref{eq:GramConstraint}, otherwise our differential equations are no longer valid. To that end, we use the momentum twistor parametrization \eqref{eq:35} and set up straight paths between two points in terms of momentum twistor variables. The starting point $\vec{s}_0$ in this parametrization is
\begin{equation}
    \vec{x}_0=\{-1,1,0,0,1,1,0,0\},
\end{equation}
and a straight line between our starting point and an endpoint is
\begin{equation}
    \vec{x}_1 (t)=(1-t)\vec{x}_0 + t \vec{x}_i.
\end{equation}

The solution at weight three is given as a one--fold integration over the weight two functions
\begin{equation}
    \vec{I}_{\text{fam}}^{\,(3)}=\vec{I}_{\text{fam}}^{\,(3)}(\vec{x}_0) + \int_0^1 \text{d} t_1 \dfrac{\text{d} A_{\text{fam}}}{\text{d}t_1} \vec{f}^{\,(2)} (t_1),
\end{equation}
where $\vec{I}_{\text{fam}}^{\,(3)}(\vec{x}_0)$ is the boundary value at weight three and $\vec{f}^{\,(2)} $ are the weight two functions expressed in terms of the momentum twistor variables. At weight four, we use integration by parts to rewrite two--fold integration over the weight--two functions as a one--fold integration
\begin{align}
    \vec{I}_{\text{fam}}^{\,(4)}&=\vec{I}_{\text{fam}}^{\,(4)}(\vec{x}_0) + \int_0^1 \text{d} t \dfrac{\text{d} A_{\text{fam}}}{\text{d}t} \vec{I}_{\text{fam}}^{\,(3)}(\vec{x}_0) +\int_0^1 \text{d} t_1 \int_0^{t_1} \text{d} t_2 \dfrac{\text{d} A_{\text{fam}}}{\text{d}t_1} \dfrac{\text{d} A_{\text{fam}}}{\text{d}t_2}  \vec{f}^{\,(2)} (t_2) \notag \\
    &=\vec{I}_{\text{fam}}^{\,(4)}(\vec{x}_0) + \int_0^1 \text{d} t \left( \dfrac{\text{d} A_{\text{fam}}}{\text{d}t} \vec{I}_{\text{fam}}^{\,(3)}(\vec{x}_0) + \left(A_{\text{fam}}(1) - A_{\text{fam}}(t) \right) \dfrac{\text{d} A_{\text{fam}}}{\text{d}t} \vec{f}^{\,(2)} (t) \right).
\end{align}

In this section, we have identified the function space for six--point Feynman integrals up to weight four in the dimensional regulator. Following the hybrid approach, the function space up to weight two is given in terms of a basis of special functions, while at weight three and four we employ a one--fold integral representation.

\subsection{Numerical validation}
\begin{table}
    \centering
    \begin{tabular}{|c|c|c|}
    \hline
        Family & {\sc AMFlow} & One--fold integration \\
        \hline
        Double--box (db) & $\sim 3.1$ h & $\sim 7.5$ min \\
        \hline
        Pentagon--trinagle (pt) & $\sim 45$ min & $\sim 1.8$ min \\
        \hline
        Hexagon--bubble (hb) & $\sim 25$ min & $\sim 1.1$ min \\
        \hline
    \end{tabular}
    \caption{Comparison of evaluation times between {\sc AMFlow}~\cite{Liu:2022chg} and a proof--of--concept implementation of one--fold integration for the three families. The precision goal for numerical one--fold integration is 20 digits. Both results were obtained on an \texttt{AMD Ryzen 7 5700G}, with 8 CPU cores.}
    \label{tab:Numerical evaluations}
\end{table}

We use the one--fold integral representation to numerically evaluate basis integrals of all three families at four different points in the bulk of the Euclidean region:
\begin{gather}
    \vec{x}_1=\left\{-\frac{299}{300},\; \frac{221}{200},\; \frac{1}{300},\; \frac{1}{200},\; \frac{53}{50}, \; \frac{21}{20},\; \frac{1}{100},\; \frac{1}{700} \right\}, \notag\\
    \vec{x}_2=\left\{ -\frac{31}{30},\; \frac{161}{150},\; \frac{43}{875},\; -\frac{17}{630},\; \frac{10153}{9450},\; \frac{243}{250},\; -\frac{22}{75},\;\frac{1}{30}\right\}, \notag \\
    \vec{x}_3=\left\{ -\dfrac{17}{10},\; \dfrac{51}{50},\; \dfrac{109}{3670},\; -\dfrac{197}{5505},\; \dfrac{3799}{3670},\; \dfrac{774}{815},\; -\frac{3}{20},\; \frac{1}{25} \right\}, \notag\\
    \vec{x}_4=\left\{ -1,\; \dfrac{9021}{8950},\; \dfrac{7}{1250},\; -\dfrac{17}{250},\; \dfrac{2003}{1790},\; \dfrac{466}{475},\; -\frac{637}{3580},\; \frac{5}{179} \right\}.
\end{gather}
Numerical values obtained via one--fold integration are in complete agreement with the numerical values obtained with {\sc AMFlow} up to the desired precision of $20$ digits. 

A proof--of--concept implementation of one--fold integration in {\sc Mathematica} is provided in the auxiliary files. This implementation is suitable for fast evaluations in the Euclidean region on the paths that do not cross any branch cuts, see Table \ref{tab:Numerical evaluations}. We leave the analytic continuation to the physical scattering region for future work. 

\section{Auxiliary files}
In this section, we provide a brief description of all auxiliary files which accompany this publication. 
\begin{itemize}
    \item \verb|MomentumTwistorParametrization.m| -- expressions for kinematic variables in terms of our momentum twistor parametrization,
    \item \verb|SqrtRepl.m| -- replacement rules for square roots $r_i$'s, $\eps_{ijkl}$'s and $\Delta_6$,
    \item \verb|Alphabet.m| -- function alphabet for the three families under study in terms of Mandelstam variables,
    \item \verb|UTBasis_fam.m| -- canonical bases for the three families,
    \item \verb|DE_fam.m| -- canonical differential equations for the three families,
    \item \verb|BoundaryConstants_fam.m| -- analytic boundary constants for all top sector and subsector integrals for all three families under study,
    \item \verb|Weight2Solutions_fam.m| -- analytic solutions at weight one and two for the three families,
    \item \verb|LineIntegration.nb| -- a proof--of--concept implementation of the one--fold integration in  {\sc Mathematica}.
\end{itemize}
\section{Summary and Outlook}
In this work, we calculated three families of genuine two--loop six--point massless Feynman integral families, double--box, pentagon--triangle and hexagon--bubble, in dimensional regularization. Building on the previous work \cite{Henn:2021cyv}, we discovered complete sets of UT integral bases for these families and provided the relevant alphabet. 
The corresponding canonical differential equations were derived via finite--field reconstruction. The boundary values for these differential equations were fixed analytically.
We solved the differential equations in the Euclidean region along several paths to weight four, in terms of classical polylogarithms and one--fold integration. The result matches the evaluation with state--of--the--art numerical packages, but provides a speed--up in evaluation time by a factor of roughly 25 even for a proof--of--concept implementation. 
This constitutes an important step towards precision computations at NNLO for $2 \to 4$ scattering with massless final states.

An important point in our method was to predict alphabet letters that may appear in the differential equations. This effectively reduces the complexity of the differential equations to knowing a set of constant quadratic matrices. This allowed us to efficiently use finite field methods for constructing those matrices. We employed an improved algorithm for the construction of odd letters that will be presented in a separate publication \cite{Matijasic:2024gkz}, which will include a dedicated computer algebra implementation.

The results for the alphabets of the three integral families we provided in this paper constitute valuable data that may be used as a benchmark for independent approaches that aim at predicting the locus of singularities of Feynman integrals, for example based on an analysis of the Landau equations \cite{Fevola:2023kaw,Fevola:2023fzn,Helmer:2024wax}, or via other methods~\cite{He:2023umf,Jiang:2024eaj}.

We also studied the property of the symbol structure of the three families. Steinmann relations for these are verified and the adjacent letters are identified. This information will be very useful for the further study of the cluster algebra structure for multileg Feynman integrals and as an input for possible bootstrap programs for six--point observables. 

Starting from the results of the present paper, there are several interesting directions for future study:
\begin{enumerate}
    \item {\it Computing the remaining 
    planar two--loop six--point Feynman integrals.} 
    The results from the present paper constitute subsector integrals that are required for computing the remaining hexagon--box, the pentagon--box and the double--pentagon integral families. 
 Canonical differential equations on the maximal for these integrals were derived in~\cite{Henn:2021cyv}.
    We expect that  following similar steps as in section~\ref{sec:PDE} it should be possible to extend their definition beyond the maximal cut, and to derive and solve the complete canonical differential equations for all these integral families. 
%
    
    \item {\it Obtaining the full six--point alphabet for two--loop amplitudes.} 
    The $223$ 
    alphabet letters identified here are likely to constitute the main part of the hexagon alphabet, but we also expect further letters to be needed for the remaining integral families.
There is evidence both from a leading singularity calculation and Schubert analysis \cite{He:2023umf} that the double--pentagon integral for $D$-dimensional external states involves elliptic integrals. For four-dimensional external states (i.e. in the dimensional reduction scheme) the answer is expected to involve only multiple polylogarithms. However, the leading singularity of that integral is rather involved, and
the corresponding odd letters \cite{Henn:2021cyv} do not seem to satisfy the factorization condition \eqref{eq:factorization}. It is therefore an open problem how to apply the algorithm we presented for predicting symbol letters to that special case.

    \item {\it Providing solutions valid in the physical region for four-jet production.}
    This is important for the application of these Feynman integrals to phenomenology. It may be achieved by transporting the known analytic boundary value to different kinematic regions. Another possibility, pursued in \cite{Chicherin:2020oor}, is to compute the boundary value for a point within the desired physical region via the physical constraints discussed here, or by numerical computation, and to then solve the canonical differential equation again in this region. 
    
\end{enumerate}

\section*{Acknowledgments}

We would like to thank S\'ergio Carr\^olo, Song He, David Kosower, Simon Telen, William J. Torres Bobadilla,  Qinglin Yang and Simone Zoia for enlightening discussions. JM gratefully acknowledges support from the Simons Center for Geometry and Physics, Stony Brook University at which some of the research for this paper was performed during the program 
``Solving N=4 super--Yang--Mills theory via Scattering Amplitudes''. This research received funding from the European Research Council (ERC) under the European Union’s Horizon 2020 research and innovation programme (grant agreement No 725110), {\it Novel structures in scattering amplitudes} and under the European Union’s Horizon Europe research and innovation programme (grant agreement No 101040760), {\it High-precision multi-leg Higgs and top physics with finite fields} (FFHiggsTop). YZ is supported from the NSF of China through Grant No. 11947301, 12047502, 12075234 and 12247103.

\appendix
\section{Identities among letters in four--dimensional kinematics}
\label{App:Identities}
In section \ref{Sec:Alphabet}, we provide an alphabet that is closed under cyclic permutations of the external legs and multiplicatively independent as functions of nine Mandelstam variables. However, in four--dimensional kinematics, the Gram determinant constraint \eqref{eq:GramConstraint} leads to a set of identities among the letters which we provide in this section.
\paragraph{Identities among rationalized letters.}
Since the $\eps_{ijkl}$ become rationalized in the four--dimensional limit, a lot of the previously algebraic letters turn into purely rational letters in terms of the momentum twistor parametrization. Here, we list a basis of 27 identities between letters in our alphabet that hold provided that the external momenta lie in a four--dimensional space.
\begin{align}
0&=T^j(-W_{141}-W_{148}+W_{208}), \quad j=0,...,5,\notag\\
0 &= T^j(-W_{142}-W_{145}+W_{196}),\quad j=0,1,2,\notag\\
0&=T^j(-W_{140}+W_{146}-W_{165}+W_{175}),\quad j=0,...,5,\notag\\
0&=T^j(-W_{141}+W_{147}+W_{153}+W_{177}+W_{184}+W_{203}),\quad j=0,1,2,\notag\\
0&=T^j(W_{153}+W_{154}+W_{177}+W_{178}+W_{198}), \quad j=0,...,4,\notag\\
0&=T^j(W_{140}-W_{146}-W_{182}-W_{184}+W_{201}-W_{203}), \quad j=0,1,\notag\\
0&=W_{144}+W_{147}+W_{155}+W_{179}-W_{187}+W_{200},\notag\\
0&=-W_{142}+W_{150}+W_{152}+W_{176}+W_{186}+W_{205}+W_{208}.
\end{align}
Again, $T$ is the generator of a cyclic permutation of the external legs, while $W_j = \log \alpha_j$. We note that the remaining permutations of the last four identities also hold, however they are not independent of the previous identities.

\paragraph{Identities among odd letters.} Our alphabet contains five square roots which are not rationalized by the momentum twistor parametrization. For the odd letters related to each of these square roots, there is one identity. In terms of the letters described in this section, the identities are given by
\begin{align}
    0&=T^j(W_{209}+W_{211}-W_{213}), \quad j=0,1,\notag\\
    0&=T^j(W_{215}-W_{218}-W_{222}),\quad j=0,1,2.
\end{align}

\bibliographystyle{JHEP}
\bibliography{sixpoint.bib}

\providecommand{\href}[2]{#2}\begingroup\raggedright\begin{thebibliography}{100}

\bibitem{ATLAS:2015xtc}
{\scshape ATLAS} collaboration, \emph{{Measurement of four-jet differential
  cross sections in $\sqrt{s}=8$ TeV proton-proton collisions using the ATLAS
  detector}}, \href{https://doi.org/10.1007/JHEP12(2015)105}{\emph{JHEP}
  {\bfseries 12} (2015) 105}
  [\href{https://arxiv.org/abs/1509.07335}{{\ttfamily 1509.07335}}].

\bibitem{Gehrmann:2018yef}
T.~Gehrmann, J.M.~Henn and N.A.~Lo~Presti, \emph{{Pentagon functions for
  massless planar scattering amplitudes}},
  \href{https://doi.org/10.1007/JHEP10(2018)103}{\emph{JHEP} {\bfseries 10}
  (2018) 103} [\href{https://arxiv.org/abs/1807.09812}{{\ttfamily
  1807.09812}}].

\bibitem{Abreu:2018rcw}
S.~Abreu, B.~Page and M.~Zeng, \emph{{Differential equations from unitarity
  cuts: nonplanar hexa-box integrals}},
  \href{https://doi.org/10.1007/JHEP01(2019)006}{\emph{JHEP} {\bfseries 01}
  (2019) 006} [\href{https://arxiv.org/abs/1807.11522}{{\ttfamily
  1807.11522}}].

\bibitem{Abreu:2018aqd}
S.~Abreu, L.J.~Dixon, E.~Herrmann, B.~Page and M.~Zeng, \emph{{The two-loop
  five-point amplitude in $\mathcal{N} =4$ super-Yang-Mills theory}},
  \href{https://doi.org/10.1103/PhysRevLett.122.121603}{\emph{Phys. Rev. Lett.}
  {\bfseries 122} (2019) 121603}
  [\href{https://arxiv.org/abs/1812.08941}{{\ttfamily 1812.08941}}].

\bibitem{Chicherin:2018mue}
D.~Chicherin, T.~Gehrmann, J.M.~Henn, N.A.~Lo~Presti, V.~Mitev and P.~Wasser,
  \emph{{Analytic result for the nonplanar hexa-box integrals}},
  \href{https://doi.org/10.1007/JHEP03(2019)042}{\emph{JHEP} {\bfseries 03}
  (2019) 042} [\href{https://arxiv.org/abs/1809.06240}{{\ttfamily
  1809.06240}}].

\bibitem{Chicherin:2018old}
D.~Chicherin, T.~Gehrmann, J.M.~Henn, P.~Wasser, Y.~Zhang and S.~Zoia,
  \emph{{All Master Integrals for Three-Jet Production at
  Next-to-Next-to-Leading Order}},
  \href{https://doi.org/10.1103/PhysRevLett.123.041603}{\emph{Phys. Rev. Lett.}
  {\bfseries 123} (2019) 041603}
  [\href{https://arxiv.org/abs/1812.11160}{{\ttfamily 1812.11160}}].

\bibitem{Chicherin:2020oor}
D.~Chicherin and V.~Sotnikov, \emph{{Pentagon Functions for Scattering of Five
  Massless Particles}},
  \href{https://doi.org/10.1007/JHEP12(2020)167}{\emph{JHEP} {\bfseries 12}
  (2020) 167} [\href{https://arxiv.org/abs/2009.07803}{{\ttfamily
  2009.07803}}].

\bibitem{Gehrmann:2015bfy}
T.~Gehrmann, J.M.~Henn and N.A.~Lo~Presti, \emph{{Analytic form of the two-loop
  planar five-gluon all-plus-helicity amplitude in QCD}},
  \href{https://doi.org/10.1103/PhysRevLett.116.062001}{\emph{Phys. Rev. Lett.}
  {\bfseries 116} (2016) 062001}
  [\href{https://arxiv.org/abs/1511.05409}{{\ttfamily 1511.05409}}].

\bibitem{Badger:2017jhb}
S.~Badger, C.~Br\o{}nnum-Hansen, H.B.~Hartanto and T.~Peraro, \emph{{First look
  at two-loop five-gluon scattering in QCD}},
  \href{https://doi.org/10.1103/PhysRevLett.120.092001}{\emph{Phys. Rev. Lett.}
  {\bfseries 120} (2018) 092001}
  [\href{https://arxiv.org/abs/1712.02229}{{\ttfamily 1712.02229}}].

\bibitem{Badger:2018enw}
S.~Badger, C.~Br\o{}nnum-Hansen, H.B.~Hartanto and T.~Peraro, \emph{{Analytic
  helicity amplitudes for two-loop five-gluon scattering: the single-minus
  case}}, \href{https://doi.org/10.1007/JHEP01(2019)186}{\emph{JHEP} {\bfseries
  01} (2019) 186} [\href{https://arxiv.org/abs/1811.11699}{{\ttfamily
  1811.11699}}].

\bibitem{Abreu:2017hqn}
S.~Abreu, F.~Febres~Cordero, H.~Ita, B.~Page and M.~Zeng, \emph{{Planar
  Two-Loop Five-Gluon Amplitudes from Numerical Unitarity}},
  \href{https://doi.org/10.1103/PhysRevD.97.116014}{\emph{Phys. Rev. D}
  {\bfseries 97} (2018) 116014}
  [\href{https://arxiv.org/abs/1712.03946}{{\ttfamily 1712.03946}}].

\bibitem{Chawdhry:2020for}
H.A.~Chawdhry, M.~Czakon, A.~Mitov and R.~Poncelet, \emph{{Two-loop
  leading-color helicity amplitudes for three-photon production at the LHC}},
  \href{https://doi.org/10.1007/JHEP06(2021)150}{\emph{JHEP} {\bfseries 06}
  (2021) 150} [\href{https://arxiv.org/abs/2012.13553}{{\ttfamily
  2012.13553}}].

\bibitem{Abreu:2020cwb}
S.~Abreu, B.~Page, E.~Pascual and V.~Sotnikov, \emph{{Leading-Color Two-Loop
  QCD Corrections for Three-Photon Production at Hadron Colliders}},
  \href{https://doi.org/10.1007/JHEP01(2021)078}{\emph{JHEP} {\bfseries 01}
  (2021) 078} [\href{https://arxiv.org/abs/2010.15834}{{\ttfamily
  2010.15834}}].

\bibitem{Abreu:2018zmy}
S.~Abreu, J.~Dormans, F.~Febres~Cordero, H.~Ita and B.~Page, \emph{{Analytic
  Form of Planar Two-Loop Five-Gluon Scattering Amplitudes in QCD}},
  \href{https://doi.org/10.1103/PhysRevLett.122.082002}{\emph{Phys. Rev. Lett.}
  {\bfseries 122} (2019) 082002}
  [\href{https://arxiv.org/abs/1812.04586}{{\ttfamily 1812.04586}}].

\bibitem{Badger:2019djh}
S.~Badger, D.~Chicherin, T.~Gehrmann, G.~Heinrich, J.~Henn, T.~Peraro et~al.,
  \emph{{Analytic form of the full two-loop five-gluon all-plus helicity
  amplitude}},
  \href{https://doi.org/10.1103/PhysRevLett.123.071601}{\emph{Phys. Rev. Lett.}
  {\bfseries 123} (2019) 071601}
  [\href{https://arxiv.org/abs/1905.03733}{{\ttfamily 1905.03733}}].

\bibitem{Abreu:2021oya}
S.~Abreu, F.F.~Cordero, H.~Ita, B.~Page and V.~Sotnikov, \emph{{Leading-color
  two-loop QCD corrections for three-jet production at hadron colliders}},
  \href{https://doi.org/10.1007/JHEP07(2021)095}{\emph{JHEP} {\bfseries 07}
  (2021) 095} [\href{https://arxiv.org/abs/2102.13609}{{\ttfamily
  2102.13609}}].

\bibitem{Agarwal:2023suw}
B.~Agarwal, F.~Buccioni, F.~Devoto, G.~Gambuti, A.~von Manteuffel and
  L.~Tancredi, \emph{{Five-Parton Scattering in QCD at Two Loops}},
  \href{https://arxiv.org/abs/2311.09870}{{\ttfamily 2311.09870}}.

\bibitem{Chawdhry:2021mkw}
H.A.~Chawdhry, M.~Czakon, A.~Mitov and R.~Poncelet, \emph{{Two-loop
  leading-colour QCD helicity amplitudes for two-photon plus jet production at
  the LHC}}, \href{https://doi.org/10.1007/JHEP07(2021)164}{\emph{JHEP}
  {\bfseries 07} (2021) 164}
  [\href{https://arxiv.org/abs/2103.04319}{{\ttfamily 2103.04319}}].

\bibitem{Agarwal:2021vdh}
B.~Agarwal, F.~Buccioni, A.~von Manteuffel and L.~Tancredi, \emph{{Two-loop
  helicity amplitudes for diphoton plus jet production in full color}},
  \href{https://arxiv.org/abs/2105.04585}{{\ttfamily 2105.04585}}.

\bibitem{DeLaurentis:2023izi}
G.~De~Laurentis, H.~Ita and V.~Sotnikov, \emph{{Double-Virtual NNLO QCD
  Corrections for Five-Parton Scattering: The Quark Channels}},
  \href{https://arxiv.org/abs/2311.18752}{{\ttfamily 2311.18752}}.

\bibitem{DeLaurentis:2023nss}
G.~De~Laurentis, H.~Ita, M.~Klinkert and V.~Sotnikov, \emph{{Double-Virtual
  NNLO QCD Corrections for Five-Parton Scattering: The Gluon Channel}},
  \href{https://arxiv.org/abs/2311.10086}{{\ttfamily 2311.10086}}.

\bibitem{Badger:2021imn}
S.~Badger, C.~Br\o{}nnum-Hansen, D.~Chicherin, T.~Gehrmann, H.B.~Hartanto,
  J.~Henn et~al., \emph{{Virtual QCD corrections to gluon-initiated diphoton
  plus jet production at hadron colliders}},
  \href{https://arxiv.org/abs/2106.08664}{{\ttfamily 2106.08664}}.

\bibitem{Kallweit:2020gcp}
S.~Kallweit, V.~Sotnikov and M.~Wiesemann, \emph{{Triphoton production at
  hadron colliders in NNLO QCD}},
  \href{https://doi.org/10.1016/j.physletb.2020.136013}{\emph{Phys. Lett. B}
  {\bfseries 812} (2021) 136013}
  [\href{https://arxiv.org/abs/2010.04681}{{\ttfamily 2010.04681}}].

\bibitem{Czakon:2021mjy}
M.~Czakon, A.~Mitov and R.~Poncelet, \emph{{Tour de force in Quantum
  Chromodynamics: A first next-to-next-to-leading order study of three-jet
  production at the LHC}},  \href{https://arxiv.org/abs/2106.05331}{{\ttfamily
  2106.05331}}.

\bibitem{Chawdhry:2021hkp}
H.A.~Chawdhry, M.~Czakon, A.~Mitov and R.~Poncelet, \emph{{NNLO QCD corrections
  to diphoton production with an additional jet at the LHC}},
  \href{https://arxiv.org/abs/2105.06940}{{\ttfamily 2105.06940}}.

\bibitem{Badger:2023mgf}
S.~Badger, M.~Czakon, H.B.~Hartanto, R.~Moodie, T.~Peraro, R.~Poncelet et~al.,
  \emph{{Isolated photon production in association with a jet pair through
  next-to-next-to-leading order in QCD}},
  \href{https://doi.org/10.1007/JHEP10(2023)071}{\emph{JHEP} {\bfseries 10}
  (2023) 071} [\href{https://arxiv.org/abs/2304.06682}{{\ttfamily
  2304.06682}}].

\bibitem{Badger:2021ohm}
S.~Badger, T.~Gehrmann, M.~Marcoli and R.~Moodie, \emph{{Next-to-leading order
  QCD corrections to diphoton-plus-jet production through gluon fusion at the
  LHC}},  \href{https://arxiv.org/abs/2109.12003}{{\ttfamily 2109.12003}}.

\bibitem{Abreu:2020jxa}
S.~Abreu, H.~Ita, F.~Moriello, B.~Page, W.~Tschernow and M.~Zeng,
  \emph{{Two-Loop Integrals for Planar Five-Point One-Mass Processes}},
  \href{https://doi.org/10.1007/JHEP11(2020)117}{\emph{JHEP} {\bfseries 11}
  (2020) 117} [\href{https://arxiv.org/abs/2005.04195}{{\ttfamily
  2005.04195}}].

\bibitem{Canko:2020ylt}
D.D.~Canko, C.G.~Papadopoulos and N.~Syrrakos, \emph{{Analytic representation
  of all planar two-loop five-point Master Integrals with one off-shell leg}},
  \href{https://doi.org/10.1007/JHEP01(2021)199}{\emph{JHEP} {\bfseries 01}
  (2021) 199} [\href{https://arxiv.org/abs/2009.13917}{{\ttfamily
  2009.13917}}].

\bibitem{Abreu:2021smk}
S.~Abreu, H.~Ita, B.~Page and W.~Tschernow, \emph{{Two-Loop Hexa-Box Integrals
  for Non-Planar Five-Point One-Mass Processes}},
  \href{https://arxiv.org/abs/2107.14180}{{\ttfamily 2107.14180}}.

\bibitem{Abreu:2023rco}
S.~Abreu, D.~Chicherin, H.~Ita, B.~Page, V.~Sotnikov, W.~Tschernow et~al.,
  \emph{{All Two-Loop Feynman Integrals for Five-Point One-Mass Scattering}},
  \href{https://arxiv.org/abs/2306.15431}{{\ttfamily 2306.15431}}.

\bibitem{Chicherin:2021dyp}
D.~Chicherin, V.~Sotnikov and S.~Zoia, \emph{{Pentagon functions for one-mass
  planar scattering amplitudes}},
  \href{https://doi.org/10.1007/JHEP01(2022)096}{\emph{JHEP} {\bfseries 01}
  (2022) 096} [\href{https://arxiv.org/abs/2110.10111}{{\ttfamily
  2110.10111}}].

\bibitem{Badger:2021nhg}
S.~Badger, H.B.~Hartanto and S.~Zoia, \emph{{Two-Loop QCD Corrections to Wbb
  Production at Hadron Colliders}},
  \href{https://doi.org/10.1103/PhysRevLett.127.012001}{\emph{Phys. Rev. Lett.}
  {\bfseries 127} (2021) 012001}
  [\href{https://arxiv.org/abs/2102.02516}{{\ttfamily 2102.02516}}].

\bibitem{Badger:2022hno}
S.~Badger, M.~Becchetti, E.~Chaubey and R.~Marzucca, \emph{{Two-loop master
  integrals for a planar topology contributing to pp \textrightarrow{}$
  t\overline{t}j $}},
  \href{https://doi.org/10.1007/JHEP01(2023)156}{\emph{JHEP} {\bfseries 01}
  (2023) 156} [\href{https://arxiv.org/abs/2210.17477}{{\ttfamily
  2210.17477}}].

\bibitem{Bern:2011ep}
Z.~Bern, G.~Diana, L.J.~Dixon, F.~Febres~Cordero, S.~Hoeche, D.A.~Kosower
  et~al., \emph{{Four-Jet Production at the Large Hadron Collider at
  Next-to-Leading Order in QCD}},
  \href{https://doi.org/10.1103/PhysRevLett.109.042001}{\emph{Phys. Rev. Lett.}
  {\bfseries 109} (2012) 042001}
  [\href{https://arxiv.org/abs/1112.3940}{{\ttfamily 1112.3940}}].

\bibitem{Badger:2013yda}
S.~Badger, B.~Biedermann, P.~Uwer and V.~Yundin, \emph{{Next-to-leading order
  QCD corrections to five jet production at the LHC}},
  \href{https://doi.org/10.1103/PhysRevD.89.034019}{\emph{Phys. Rev. D}
  {\bfseries 89} (2014) 034019}
  [\href{https://arxiv.org/abs/1309.6585}{{\ttfamily 1309.6585}}].

\bibitem{Ellis:2006ss}
R.K.~Ellis, W.T.~Giele and G.~Zanderighi, \emph{{The One-loop amplitude for
  six-gluon scattering}},
  \href{https://doi.org/10.1088/1126-6708/2006/05/027}{\emph{JHEP} {\bfseries
  05} (2006) 027} [\href{https://arxiv.org/abs/hep-ph/0602185}{{\ttfamily
  hep-ph/0602185}}].

\bibitem{Dunbar:2008zz}
D.C.~Dunbar, \emph{{The six gluon one-loop amplitude}},
  \href{https://doi.org/10.1016/j.nuclphysbps.2008.09.093}{\emph{Nucl. Phys. B
  Proc. Suppl.} {\bfseries 183} (2008) 122}.

\bibitem{PhysRevD.49.2197}
G.~Mahlon, \emph{One loop multiphoton helicity amplitudes},
  \href{https://doi.org/10.1103/PhysRevD.49.2197}{\emph{Phys. Rev. D}
  {\bfseries 49} (1994) 2197}.

\bibitem{Ossola:2007bb}
G.~Ossola, C.G.~Papadopoulos and R.~Pittau, \emph{{Numerical evaluation of
  six-photon amplitudes}},
  \href{https://doi.org/10.1088/1126-6708/2007/07/085}{\emph{JHEP} {\bfseries
  07} (2007) 085} [\href{https://arxiv.org/abs/0704.1271}{{\ttfamily
  0704.1271}}].

\bibitem{Binoth:2007ca}
T.~Binoth, G.~Heinrich, T.~Gehrmann and P.~Mastrolia, \emph{{Six-Photon
  Amplitudes}},
  \href{https://doi.org/10.1016/j.physletb.2007.04.032}{\emph{Phys. Lett. B}
  {\bfseries 649} (2007) 422}
  [\href{https://arxiv.org/abs/hep-ph/0703311}{{\ttfamily hep-ph/0703311}}].

\bibitem{Bernicot:2007hs}
C.~Bernicot and J.P.~Guillet, \emph{{Six-Photon Amplitudes in Scalar QED}},
  \href{https://doi.org/10.1088/1126-6708/2008/01/059}{\emph{JHEP} {\bfseries
  01} (2008) 059} [\href{https://arxiv.org/abs/0711.4713}{{\ttfamily
  0711.4713}}].

\bibitem{Dunbar:2016gjb}
D.C.~Dunbar, G.R.~Jehu and W.B.~Perkins, \emph{{Two-loop six gluon all plus
  helicity amplitude}},
  \href{https://doi.org/10.1103/PhysRevLett.117.061602}{\emph{Phys. Rev. Lett.}
  {\bfseries 117} (2016) 061602}
  [\href{https://arxiv.org/abs/1605.06351}{{\ttfamily 1605.06351}}].

\bibitem{Dunbar:2017nfy}
D.C.~Dunbar, J.H.~Godwin, G.R.~Jehu and W.B.~Perkins, \emph{{Analytic
  all-plus-helicity gluon amplitudes in QCD}},
  \href{https://doi.org/10.1103/PhysRevD.96.116013}{\emph{Phys. Rev. D}
  {\bfseries 96} (2017) 116013}
  [\href{https://arxiv.org/abs/1710.10071}{{\ttfamily 1710.10071}}].

\bibitem{Golden:2013xva}
J.~Golden, A.B.~Goncharov, M.~Spradlin, C.~Vergu and A.~Volovich,
  \emph{{Motivic Amplitudes and Cluster Coordinates}},
  \href{https://doi.org/10.1007/JHEP01(2014)091}{\emph{JHEP} {\bfseries 01}
  (2014) 091} [\href{https://arxiv.org/abs/1305.1617}{{\ttfamily 1305.1617}}].

\bibitem{Caron-Huot:2016owq}
S.~Caron-Huot, L.J.~Dixon, A.~McLeod and M.~von Hippel, \emph{{Bootstrapping a
  Five-Loop Amplitude Using Steinmann Relations}},
  \href{https://doi.org/10.1103/PhysRevLett.117.241601}{\emph{Phys. Rev. Lett.}
  {\bfseries 117} (2016) 241601}
  [\href{https://arxiv.org/abs/1609.00669}{{\ttfamily 1609.00669}}].

\bibitem{Dixon:2023kop}
L.J.~Dixon and Y.-T.~Liu, \emph{{An eight loop amplitude via antipodal
  duality}}, \href{https://doi.org/10.1007/JHEP09(2023)098}{\emph{JHEP}
  {\bfseries 09} (2023) 098}
  [\href{https://arxiv.org/abs/2308.08199}{{\ttfamily 2308.08199}}].

\bibitem{Drummond:2008vq}
J.M.~Drummond, J.~Henn, G.P.~Korchemsky and E.~Sokatchev, \emph{{Dual
  superconformal symmetry of scattering amplitudes in N=4 super-Yang-Mills
  theory}}, \href{https://doi.org/10.1016/j.nuclphysb.2009.11.022}{\emph{Nucl.
  Phys. B} {\bfseries 828} (2010) 317}
  [\href{https://arxiv.org/abs/0807.1095}{{\ttfamily 0807.1095}}].

\bibitem{Chicherin:2020umh}
D.~Chicherin, J.M.~Henn and G.~Papathanasiou, \emph{{Cluster algebras for
  Feynman integrals}},
  \href{https://doi.org/10.1103/PhysRevLett.126.091603}{\emph{Phys. Rev. Lett.}
  {\bfseries 126} (2021) 091603}
  [\href{https://arxiv.org/abs/2012.12285}{{\ttfamily 2012.12285}}].

\bibitem{Bossinger:2022eiy}
L.~Bossinger, J.M.~Drummond and R.~Glew, \emph{{Adjacency for scattering
  amplitudes from the Gr\"obner fan}},
  \href{https://doi.org/10.1007/JHEP11(2023)002}{\emph{JHEP} {\bfseries 11}
  (2023) 002} [\href{https://arxiv.org/abs/2212.08931}{{\ttfamily
  2212.08931}}].

\bibitem{Arkani-Hamed:2012zlh}
N.~Arkani-Hamed, J.L.~Bourjaily, F.~Cachazo, A.B.~Goncharov, A.~Postnikov and
  J.~Trnka, \emph{{Grassmannian Geometry of Scattering Amplitudes}}, Cambridge
  University Press (4, 2016),
  \href{https://doi.org/10.1017/CBO9781316091548}{10.1017/CBO9781316091548},
  [\href{https://arxiv.org/abs/1212.5605}{{\ttfamily 1212.5605}}].

\bibitem{Arkani-Hamed:2019rds}
N.~Arkani-Hamed, T.~Lam and M.~Spradlin, \emph{{Non-perturbative geometries for
  planar $ \mathcal{N} $ = 4 SYM amplitudes}},
  \href{https://doi.org/10.1007/JHEP03(2021)065}{\emph{JHEP} {\bfseries 03}
  (2021) 065} [\href{https://arxiv.org/abs/1912.08222}{{\ttfamily
  1912.08222}}].

\bibitem{Herderschee:2021dez}
A.~Herderschee, \emph{{Algebraic branch points at all loop orders from positive
  kinematics and wall crossing}},
  \href{https://doi.org/10.1007/JHEP07(2021)049}{\emph{JHEP} {\bfseries 07}
  (2021) 049} [\href{https://arxiv.org/abs/2102.03611}{{\ttfamily
  2102.03611}}].

\bibitem{Ren:2021ztg}
L.~Ren, M.~Spradlin and A.~Volovich, \emph{{Symbol alphabets from tensor
  diagrams}}, \href{https://doi.org/10.1007/JHEP12(2021)079}{\emph{JHEP}
  {\bfseries 12} (2021) 079}
  [\href{https://arxiv.org/abs/2106.01405}{{\ttfamily 2106.01405}}].

\bibitem{Henke:2021ity}
N.~Henke and G.~Papathanasiou, \emph{{Singularities of eight- and nine-particle
  amplitudes from cluster algebras and tropical geometry}},
  \href{https://doi.org/10.1007/JHEP10(2021)007}{\emph{JHEP} {\bfseries 10}
  (2021) 007} [\href{https://arxiv.org/abs/2106.01392}{{\ttfamily
  2106.01392}}].

\bibitem{He:2021non}
S.~He, Z.~Li and Q.~Yang, \emph{{Truncated cluster algebras and Feynman
  integrals with algebraic letters}},
  \href{https://doi.org/10.1007/JHEP12(2021)110}{\emph{JHEP} {\bfseries 12}
  (2021) 110} [\href{https://arxiv.org/abs/2106.09314}{{\ttfamily
  2106.09314}}].

\bibitem{Yang:2022gko}
Q.~Yang, \emph{{Schubert problems, positivity and symbol letters}},
  \href{https://doi.org/10.1007/JHEP08(2022)168}{\emph{JHEP} {\bfseries 08}
  (2022) 168} [\href{https://arxiv.org/abs/2203.16112}{{\ttfamily
  2203.16112}}].

\bibitem{He:2022ujv}
S.~He, Z.~Li and C.~Zhang, \emph{{A nice two-loop next-to-next-to-MHV amplitude
  in $ \mathcal{N} $ = 4 super-Yang-Mills}},
  \href{https://doi.org/10.1007/JHEP12(2022)158}{\emph{JHEP} {\bfseries 12}
  (2022) 158} [\href{https://arxiv.org/abs/2209.10856}{{\ttfamily
  2209.10856}}].

\bibitem{Fevola:2023kaw}
C.~Fevola, S.~Mizera and S.~Telen, \emph{{Landau Singularities Revisited:
  Computational Algebraic Geometry for Feynman Integrals}},
  \href{https://doi.org/10.1103/PhysRevLett.132.101601}{\emph{Phys. Rev. Lett.}
  {\bfseries 132} (2024) 101601}
  [\href{https://arxiv.org/abs/2311.14669}{{\ttfamily 2311.14669}}].

\bibitem{Fevola:2023fzn}
C.~Fevola, S.~Mizera and S.~Telen, \emph{{Principal Landau Determinants}},
  \href{https://arxiv.org/abs/2311.16219}{{\ttfamily 2311.16219}}.

\bibitem{Helmer:2024wax}
M.~Helmer, G.~Papathanasiou and F.~Tellander, \emph{{Landau Singularities from
  Whitney Stratifications}},
  \href{https://arxiv.org/abs/2402.14787}{{\ttfamily 2402.14787}}.

\bibitem{He:2023umf}
S.~He, X.~Jiang, J.~Liu and Q.~Yang, \emph{{On symbology and differential
  equations of Feynman integrals from Schubert analysis}},
  \href{https://doi.org/10.1007/JHEP12(2023)140}{\emph{JHEP} {\bfseries 12}
  (2023) 140} [\href{https://arxiv.org/abs/2309.16441}{{\ttfamily
  2309.16441}}].

\bibitem{Jiang:2024eaj}
X.~Jiang, J.~Liu, X.~Xu and L.L.~Yang, \emph{{Symbol letters of Feynman
  integrals from Gram determinants}},
  \href{https://arxiv.org/abs/2401.07632}{{\ttfamily 2401.07632}}.

\bibitem{Alday:2011ga}
L.F.~Alday, E.I.~Buchbinder and A.A.~Tseytlin, \emph{{Correlation function of
  null polygonal Wilson loops with local operators}},
  \href{https://doi.org/10.1007/JHEP09(2011)034}{\emph{JHEP} {\bfseries 09}
  (2011) 034} [\href{https://arxiv.org/abs/1107.5702}{{\ttfamily 1107.5702}}].

\bibitem{Chicherin:2022bov}
D.~Chicherin and J.M.~Henn, \emph{{Symmetry properties of Wilson loops with a
  Lagrangian insertion}},
  \href{https://doi.org/10.1007/JHEP07(2022)057}{\emph{JHEP} {\bfseries 07}
  (2022) 057} [\href{https://arxiv.org/abs/2202.05596}{{\ttfamily
  2202.05596}}].

\bibitem{Chicherin:2022zxo}
D.~Chicherin and J.~Henn, \emph{{Pentagon Wilson loop with Lagrangian insertion
  at two loops in $ \mathcal{N} $ = 4 super Yang-Mills theory}},
  \href{https://doi.org/10.1007/JHEP07(2022)038}{\emph{JHEP} {\bfseries 07}
  (2022) 038} [\href{https://arxiv.org/abs/2204.00329}{{\ttfamily
  2204.00329}}].

\bibitem{Henn:2021cyv}
J.~Henn, T.~Peraro, Y.~Xu and Y.~Zhang, \emph{{A first look at the function
  space for planar two-loop six-particle Feynman integrals}},
  \href{https://doi.org/10.1007/JHEP03(2022)056}{\emph{JHEP} {\bfseries 03}
  (2022) 056} [\href{https://arxiv.org/abs/2112.10605}{{\ttfamily
  2112.10605}}].

\bibitem{Henn:2013pwa}
J.M.~Henn, \emph{{Multiloop integrals in dimensional regularization made
  simple}}, \href{https://doi.org/10.1103/PhysRevLett.110.251601}{\emph{Phys.
  Rev. Lett.} {\bfseries 110} (2013) 251601}
  [\href{https://arxiv.org/abs/1304.1806}{{\ttfamily 1304.1806}}].

\bibitem{Henn:2022ydo}
J.M.~Henn, A.~Matija\v{s}i\'c and J.~Miczajka, \emph{{One-loop hexagon integral
  to higher orders in the dimensional regulator}},
  \href{https://doi.org/10.1007/JHEP01(2023)096}{\emph{JHEP} {\bfseries 01}
  (2023) 096} [\href{https://arxiv.org/abs/2210.13505}{{\ttfamily
  2210.13505}}].

\bibitem{Baikov:1996cd}
P.A.~Baikov, \emph{{Explicit solutions of n loop vacuum integral recurrence
  relations}},  \href{https://arxiv.org/abs/hep-ph/9604254}{{\ttfamily
  hep-ph/9604254}}.

\bibitem{Dlapa:2021qsl}
C.~Dlapa, X.~Li and Y.~Zhang, \emph{{Leading singularities in Baikov
  representation and Feynman integrals with uniform transcendental weight}},
  \href{https://arxiv.org/abs/2103.04638}{{\ttfamily 2103.04638}}.

\bibitem{Chen:2022lzr}
J.~Chen, X.~Jiang, C.~Ma, X.~Xu and L.L.~Yang, \emph{{Baikov representations,
  intersection theory, and canonical Feynman integrals}},
  \href{https://doi.org/10.1007/JHEP07(2022)066}{\emph{JHEP} {\bfseries 07}
  (2022) 066} [\href{https://arxiv.org/abs/2202.08127}{{\ttfamily
  2202.08127}}].

\bibitem{vonManteuffel:2014ixa}
A.~von Manteuffel and R.M.~Schabinger, \emph{{A novel approach to integration
  by parts reduction}},
  \href{https://doi.org/10.1016/j.physletb.2015.03.029}{\emph{Phys. Lett. B}
  {\bfseries 744} (2015) 101}
  [\href{https://arxiv.org/abs/1406.4513}{{\ttfamily 1406.4513}}].

\bibitem{Peraro:2016wsq}
T.~Peraro, \emph{{Scattering amplitudes over finite fields and multivariate
  functional reconstruction}},
  \href{https://doi.org/10.1007/JHEP12(2016)030}{\emph{JHEP} {\bfseries 12}
  (2016) 030} [\href{https://arxiv.org/abs/1608.01902}{{\ttfamily
  1608.01902}}].

\bibitem{Matijasic:2024gkz}
A.~Matija\v{s}i\'{c} and J.~Miczajka, \emph{{Effortless: Efficient generation
  of odd letters with multiple roots as leading singularities}}, In preparation
  (2024).

\bibitem{Hodges:2009hk}
A.~Hodges, \emph{{Eliminating spurious poles from gauge-theoretic amplitudes}},
  \href{https://doi.org/10.1007/JHEP05(2013)135}{\emph{JHEP} {\bfseries 05}
  (2013) 135} [\href{https://arxiv.org/abs/0905.1473}{{\ttfamily 0905.1473}}].

\bibitem{Eden:1966dnq}
R.J.~Eden, P.V.~Landshoff, D.I.~Olive and J.C.~Polkinghorne, \emph{{The
  analytic S-matrix}}, Cambridge Univ. Press, Cambridge (1966).

\bibitem{Byckling:1971vca}
E.~Byckling and K.~Kajantie, \emph{{Particle Kinematics}: {(Chapters I-VI,
  X)}}, University of Jyvaskyla, Jyvaskyla, Finland (1971).

\bibitem{Badger:2013gxa}
S.~Badger, H.~Frellesvig and Y.~Zhang, \emph{{A Two-Loop Five-Gluon Helicity
  Amplitude in QCD}},
  \href{https://doi.org/10.1007/JHEP12(2013)045}{\emph{JHEP} {\bfseries 12}
  (2013) 045} [\href{https://arxiv.org/abs/1310.1051}{{\ttfamily 1310.1051}}].

\bibitem{Chetyrkin:1981qh}
K.G.~Chetyrkin and F.V.~Tkachov, \emph{{Integration by Parts: The Algorithm to
  Calculate beta Functions in 4 Loops}},
  \href{https://doi.org/10.1016/0550-3213(81)90199-1}{\emph{Nucl. Phys.}
  {\bfseries B192} (1981) 159}.

\bibitem{Laporta:2000dsw}
S.~Laporta, \emph{{High precision calculation of multiloop Feynman integrals by
  difference equations}},
  \href{https://doi.org/10.1142/S0217751X00002159}{\emph{Int. J. Mod. Phys. A}
  {\bfseries 15} (2000) 5087}
  [\href{https://arxiv.org/abs/hep-ph/0102033}{{\ttfamily hep-ph/0102033}}].

\bibitem{Lee:2012cn}
R.N.~Lee, \emph{{Presenting LiteRed: a tool for the Loop InTEgrals REDuction}},
   \href{https://arxiv.org/abs/1212.2685}{{\ttfamily 1212.2685}}.

\bibitem{Heinrich:2008si}
G.~Heinrich, \emph{{Sector Decomposition}},
  \href{https://doi.org/10.1142/S0217751X08040263}{\emph{Int. J. Mod. Phys. A}
  {\bfseries 23} (2008) 1457}
  [\href{https://arxiv.org/abs/0803.4177}{{\ttfamily 0803.4177}}].

\bibitem{Heinrich:2023til}
G.~Heinrich, S.P.~Jones, M.~Kerner, V.~Magerya, A.~Olsson and J.~Schlenk,
  \emph{{Numerical scattering amplitudes with pySecDec}},
  \href{https://doi.org/10.1016/j.cpc.2023.108956}{\emph{Comput. Phys. Commun.}
  {\bfseries 295} (2024) 108956}
  [\href{https://arxiv.org/abs/2305.19768}{{\ttfamily 2305.19768}}].

\bibitem{Liu:2022chg}
X.~Liu and Y.-Q.~Ma, \emph{{AMFlow: A Mathematica package for Feynman integrals
  computation via auxiliary mass flow}},
  \href{https://doi.org/10.1016/j.cpc.2022.108565}{\emph{Comput. Phys. Commun.}
  {\bfseries 283} (2023) 108565}
  [\href{https://arxiv.org/abs/2201.11669}{{\ttfamily 2201.11669}}].

\bibitem{Kotikov:1990kg}
A.~Kotikov, \emph{{Differential equations method: New technique for massive
  Feynman diagrams calculation}},
  \href{https://doi.org/10.1016/0370-2693(91)90413-K}{\emph{Phys. Lett. B}
  {\bfseries 254} (1991) 158}.

\bibitem{Remiddi:1997ny}
E.~Remiddi, \emph{{Differential equations for Feynman graph amplitudes}},
  {\emph{Nuovo Cim. A} {\bfseries 110} (1997) 1435}
  [\href{https://arxiv.org/abs/hep-th/9711188}{{\ttfamily hep-th/9711188}}].

\bibitem{Peraro:2019svx}
T.~Peraro, \emph{{FiniteFlow: multivariate functional reconstruction using
  finite fields and dataflow graphs}},
  \href{https://doi.org/10.1007/JHEP07(2019)031}{\emph{JHEP} {\bfseries 07}
  (2019) 031} [\href{https://arxiv.org/abs/1905.08019}{{\ttfamily
  1905.08019}}].

\bibitem{Arkani-Hamed:2010pyv}
N.~Arkani-Hamed, J.L.~Bourjaily, F.~Cachazo and J.~Trnka, \emph{{Local
  Integrals for Planar Scattering Amplitudes}},
  \href{https://doi.org/10.1007/JHEP06(2012)125}{\emph{JHEP} {\bfseries 06}
  (2012) 125} [\href{https://arxiv.org/abs/1012.6032}{{\ttfamily 1012.6032}}].

\bibitem{Bjorken:1959fd}
J.D.~Bjorken, \emph{{Experimental tests of Quantum electrodynamics and spectral
  representations of Green's functions in perturbation theory}}, Ph.D. thesis,
  Stanford U., 1959.

\bibitem{Landau:1959fi}
L.D.~Landau, \emph{{On analytic properties of vertex parts in quantum field
  theory}},
  \href{https://doi.org/10.1016/B978-0-08-010586-4.50103-6}{\emph{Nucl. Phys.}
  {\bfseries 13} (1959) 181}.

\bibitem{Nakanishi:10.1143/PTP.22.128}
N.~Nakanishi, \emph{{Ordinary and Anomalous Thresholds in Perturbation
  Theory}}, \href{https://doi.org/10.1143/PTP.22.128}{\emph{Progress of
  Theoretical Physics} {\bfseries 22} (1959) 128}
  [\href{https://arxiv.org/abs/https://academic.oup.com/ptp/article-pdf/22/1/128/5427385/22-1-128.pdf}{{\ttfamily
  https://academic.oup.com/ptp/article-pdf/22/1/128/5427385/22-1-128.pdf}}].

\bibitem{Heller:2019gkq}
M.~Heller, A.~von Manteuffel and R.M.~Schabinger, \emph{{Multiple
  polylogarithms with algebraic arguments and the two-loop EW-QCD Drell-Yan
  master integrals}},
  \href{https://doi.org/10.1103/PhysRevD.102.016025}{\emph{Phys. Rev. D}
  {\bfseries 102} (2020) 016025}
  [\href{https://arxiv.org/abs/1907.00491}{{\ttfamily 1907.00491}}].

\bibitem{Zoia:2021zmb}
S.~Zoia, \emph{{Modern analytic methods for scattering amplitudes, with an
  application to two-loop five-particle processes}}, Ph.D. thesis,
  Ludwig-Maximilians-Universitat Munchen, 9, 2021.

\bibitem{FebresCordero:2023gjh}
F.~Febres~Cordero, G.~Figueiredo, M.~Kraus, B.~Page and L.~Reina,
  \emph{{Two-Loop Master Integrals for Leading-Color $pp\to t\bar{t}H$
  Amplitudes with a Light-Quark Loop}},
  \href{https://arxiv.org/abs/2312.08131}{{\ttfamily 2312.08131}}.

\bibitem{Dixon:2011ng}
L.J.~Dixon, J.M.~Drummond and J.M.~Henn, \emph{{The one-loop six-dimensional
  hexagon integral and its relation to MHV amplitudes in N=4 SYM}},
  \href{https://doi.org/10.1007/JHEP06(2011)100}{\emph{JHEP} {\bfseries 06}
  (2011) 100} [\href{https://arxiv.org/abs/1104.2787}{{\ttfamily 1104.2787}}].

\bibitem{Dixon:2011pw}
L.J.~Dixon, J.M.~Drummond and J.M.~Henn, \emph{{Bootstrapping the three-loop
  hexagon}}, \href{https://doi.org/10.1007/JHEP11(2011)023}{\emph{JHEP}
  {\bfseries 11} (2011) 023} [\href{https://arxiv.org/abs/1108.4461}{{\ttfamily
  1108.4461}}].

\bibitem{Henn:2013nsa}
J.M.~Henn, A.V.~Smirnov and V.A.~Smirnov, \emph{{Evaluating single-scale and/or
  non-planar diagrams by differential equations}},
  \href{https://doi.org/10.1007/JHEP03(2014)088}{\emph{JHEP} {\bfseries 03}
  (2014) 088} [\href{https://arxiv.org/abs/1312.2588}{{\ttfamily 1312.2588}}].

\bibitem{Henn:2020lye}
J.~Henn, B.~Mistlberger, V.A.~Smirnov and P.~Wasser, \emph{{Constructing d-log
  integrands and computing master integrals for three-loop four-particle
  scattering}}, \href{https://doi.org/10.1007/JHEP04(2020)167}{\emph{JHEP}
  {\bfseries 04} (2020) 167}
  [\href{https://arxiv.org/abs/2002.09492}{{\ttfamily 2002.09492}}].

\bibitem{Goncharov:1998kja}
A.B.~Goncharov, \emph{{Multiple polylogarithms, cyclotomy and modular
  complexes}}, \href{https://doi.org/10.4310/MRL.1998.v5.n4.a7}{\emph{Math.
  Res. Lett.} {\bfseries 5} (1998) 497}
  [\href{https://arxiv.org/abs/1105.2076}{{\ttfamily 1105.2076}}].

\bibitem{Henn:2015sem}
J.M.~Henn, A.V.~Smirnov and V.A.~Smirnov, \emph{{Evaluating Multiple
  Polylogarithm Values at Sixth Roots of Unity up to Weight Six}},
  \href{https://doi.org/10.1016/j.nuclphysb.2017.03.026}{\emph{Nucl. Phys. B}
  {\bfseries 919} (2017) 315}
  [\href{https://arxiv.org/abs/1512.08389}{{\ttfamily 1512.08389}}].

\bibitem{Duhr:2019tlz}
C.~Duhr and F.~Dulat, \emph{{PolyLogTools \textemdash{} polylogs for the
  masses}}, \href{https://doi.org/10.1007/JHEP08(2019)135}{\emph{JHEP}
  {\bfseries 08} (2019) 135}
  [\href{https://arxiv.org/abs/1904.07279}{{\ttfamily 1904.07279}}].

\bibitem{Chen:1977oja}
K.-T.~Chen, \emph{{Iterated path integrals}},
  \href{https://doi.org/10.1090/S0002-9904-1977-14320-6}{\emph{Bull. Am. Math.
  Soc.} {\bfseries 83} (1977) 831}.

\bibitem{Goncharov:2010jf}
A.B.~Goncharov, M.~Spradlin, C.~Vergu and A.~Volovich, \emph{{Classical
  Polylogarithms for Amplitudes and Wilson Loops}},
  \href{https://doi.org/10.1103/PhysRevLett.105.151605}{\emph{Phys.\ Rev.\
  Lett.} {\bfseries 105} (2010) 151605}
  [\href{https://arxiv.org/abs/1006.5703}{{\ttfamily 1006.5703}}].

\bibitem{Caron-Huot:2019bsq}
S.~Caron-Huot, L.J.~Dixon, F.~Dulat, M.~Von~Hippel, A.J.~McLeod and
  G.~Papathanasiou, \emph{{The Cosmic Galois Group and Extended Steinmann
  Relations for Planar $\mathcal{N} = 4$ SYM Amplitudes}},
  \href{https://doi.org/10.1007/JHEP09(2019)061}{\emph{JHEP} {\bfseries 09}
  (2019) 061} [\href{https://arxiv.org/abs/1906.07116}{{\ttfamily
  1906.07116}}].

\bibitem{Steinmann:1960a}
O.~Steinmann, \emph{{\"{U}ber den Zusammenhang zwischen den Wightmanfunktionen
  und den retardierten Kommutatoren}},
  \href{https://doi.org/10.5169/seals-113076}{\emph{Helv. Physica Acta}
  {\bfseries 33} (1960) 257}.

\bibitem{Steinmann:1960b}
O.~Steinmann, \emph{{Wightman-Funktionen und retardierte Kommutatoren. II}},
  \href{https://doi.org/10.5169/seals-113079}{\emph{Helv. Physica Acta}
  {\bfseries 33} (1960) 347}.

\bibitem{Dixon:2011nj}
L.J.~Dixon, J.M.~Drummond and J.M.~Henn, \emph{{Analytic result for the
  two-loop six-point NMHV amplitude in N=4 super Yang-Mills theory}},
  \href{https://doi.org/10.1007/JHEP01(2012)024}{\emph{JHEP} {\bfseries 01}
  (2012) 024} [\href{https://arxiv.org/abs/1111.1704}{{\ttfamily 1111.1704}}].

\bibitem{Caron-Huot:2020bkp}
S.~Caron-Huot, L.J.~Dixon, J.M.~Drummond, F.~Dulat, J.~Foster,
  O.~G\"urdo\u{g}an et~al., \emph{{The Steinmann Cluster Bootstrap for $N$ = 4
  Super Yang-Mills Amplitudes}},
  \href{https://doi.org/10.22323/1.376.0003}{\emph{PoS} {\bfseries CORFU2019}
  (2020) 003} [\href{https://arxiv.org/abs/2005.06735}{{\ttfamily
  2005.06735}}].

\bibitem{Hannesdottir:2021kpd}
H.S.~Hannesdottir, A.J.~McLeod, M.D.~Schwartz and C.~Vergu, \emph{{Implications
  of the Landau equations for iterated integrals}},
  \href{https://doi.org/10.1103/PhysRevD.105.L061701}{\emph{Phys. Rev. D}
  {\bfseries 105} (2022) L061701}
  [\href{https://arxiv.org/abs/2109.09744}{{\ttfamily 2109.09744}}].

\bibitem{Caron-Huot:2014lda}
S.~Caron-Huot and J.M.~Henn, \emph{{Iterative structure of finite loop
  integrals}}, \href{https://doi.org/10.1007/JHEP06(2014)114}{\emph{JHEP}
  {\bfseries 06} (2014) 114} [\href{https://arxiv.org/abs/1404.2922}{{\ttfamily
  1404.2922}}].

\end{thebibliography}\endgroup
\end{document}